\documentclass[acmsmall]{acmart}
\usepackage{comment}
\usepackage[linesnumbered,ruled,vlined]{algorithm2e}
\usepackage{algpseudocode}  
\usepackage{textgreek}
\usepackage{tablefootnote}
\usepackage{multirow}
\usepackage{overpic}
\usepackage{subfigure}
\usepackage{enumitem}
\usepackage{bm}
\usepackage{xcolor}
\usepackage{tabu}

\newcommand\ourmethod{\textsc{RusKey}}
\newcommand\RLmodel{\textsc{Lerp}}
\newcommand\FLSM{\textsc{FLSM-tree}}
\newcommand\Monkey{Monkey}
\newcommand\tuner{{tuning model}}
\newcommand\network{actor-critic network}
\newcommand\aggr{\textsc{Aggressive}}
\newcommand\lazy{\textsc{Lazy}}
\newcommand\mode{\textsc{Moderate}}
\newcommand\lazyl{\textsc{Lazy-Leveling}}

\setcopyright{acmcopyright}
\copyrightyear{2018}
\acmYear{2018}
\acmDOI{XXXXXXX.XXXXXXX}

\acmConference [SIGMOD' 24]{[Proceedings of the 2024 International Conference on Management of Data}{June 11--16,
  2024}{San, NY}

\acmPrice{15.00}
\acmISBN{978-1-4503-XXXX-X/18/06}

\begin{document}
\title{Learning to Optimize LSM-trees: Towards A Reinforcement\\ Learning based Key-Value Store for Dynamic Workloads}

\author{Dingheng Mo}
\affiliation{%
  \institution{Nanyang Technological University}
  \city{}
  \country{Singapore}}
\email{}

\author{Fanchao Chen}
\authornote{Work done when the author was working as a research assistant under the supervision of Siqiang Luo.}
\affiliation{%
  \institution{Fudan University}
  \city{Shanghai}
  \country{China}}
\email{}

\author{Siqiang Luo}
\authornote{Siqiang Luo is the corresponding author.}
\affiliation{%
  \institution{Nanyang Technological University}
  \city{}
  \country{Singapore}}
\email{}

\author{Caihua Shan}
\affiliation{%
  \institution{Microsoft Research Asia}
  \city{Beijing}
  \country{China}}
\email{}

\begin{abstract}
LSM-trees are widely adopted as the storage backends of key-value stores. However, optimizing the system performance under dynamic workloads has not been sufficiently studied in previous work. 
To fill the gap, we present {\ourmethod}, 
a key-value store with the following new features: 
(1) {\ourmethod} is a first attempt to design LSM-tree structures online to enable robust performance under the context of dynamic workloads; (2) {\ourmethod} is the first study to use Reinforcement Learning (RL) to guide LSM-tree transformations; (3) {\ourmethod} includes a new LSM-tree design, named {\FLSM}, that facilitates efficient transitions between different compaction policies, which addresses the key bottleneck for dynamic key-value stores. We justify the superiority of the new design with theoretical analysis; (4) {\ourmethod} requires no prior workload knowledge for system adjustment, in contrast to state-of-the-art techniques. Experiments show that {\ourmethod} exhibits strong performance robustness across diverse workloads, achieving up to 4x better end-to-end performance than the RocksDB system under various settings. 
\end{abstract}

\maketitle

\section{Introduction}\label{sec:intro}

Log-structured-merge-trees (LSM-trees) are widely adopted as the backbones of mainstream key-value stores, such as LevelDB \cite{leveldb} at
Google, RocksDB \cite{rocksdb} at Facebook, Cassandra \cite{lakshman2010cassandra} at Apache, Voldemort \cite{tom} at LinkedIn, and WiredTiger \cite{wiredtiger} at MongoDB. An LSM-tree stores data as key-value entries, offering efficient entry updates and key lookups. It organizes the data into multiple levels with exponentially increasing capacities.
The data in each level is organized into sorted {\it runs}. 
An LSM-tree appends the new key-value entry into the main memory buffer, whose filled-up will sort the buffered entries and compact them as a larger sorted run merged into the next level. The process can cascade down whenever the affected level reaches its capacity.   
The compaction policy in a level determines how frequently the level compacts its data.

In LSM-tree based key-value stores, one important topic is to handle dynamically changing workloads. As demonstrated by prior studies (e.g., ~\cite{atikoglu2012workload, ma2018query, curino2011workload, gmach2007workload}), workloads for
real-world applications are dynamically changing with daily patterns and shifts, which underscores the necessity of processing various portions of key lookups and entry updates. For example, Cao {\it et al.}~\cite{cao2020characterizing} collected a trace of 14-day operations processed by UDB~\cite{armstrong2013linkbench}, which is a database storing Facebook's social graph data. They concluded that the numbers and portions of queries and updates vary significantly over time.  This is not surprising as modern databases have billions of users who are actively accessing or updating database applications in different time spans. 
Therefore, in this paper we focus on designing an LSM-tree key-value store that can process dynamic workloads with high performance.

\subsection{Prior Work}
There has been fast development on LSM-tree optimizations. Behind the ingenuity of an optimized LSM-tree often we see an elaborate
design of its key components. Examples include optimizing compaction policies \cite{dostoevsky2018, idreos2019designcontinuum, chatterjee2021cosine, huynh2021endure, dayan2019log, sarkar2022constructing},
developing update-friendly compaction schemes \cite{sears2012blsm, dayanspooky, sarkar2020lethe, alkowaileet2019lsm, wu2015lsmtrie, yao2017light, yao2017building},
enhancing Bloom filters \cite{dayan2021chucky, zhang2018elasticbf, zhu2021reducing, dayan2017monkey, qader2018comparative, li2022seesaw},
enhancing range filters~\cite{luo2020rosetta,knorr2022proteus,zhang2018surf},
adopting unconventional hardwares \cite{ahmad2015compaction, huang2019x, vinccon2018noftl, zhang2020fpga, wang2014efficient, thonangi2017log}, 
{
\color{black}
narrowing the gap between LSM-trees and
update-in-place designs
\cite{yu2022treeline},
}
separating keys and values \cite{chan2018hashkv, lu2017wisckey},
keeping hot entries in the buffer and selective flushing \cite{balmau2017triad},
improving concurrency \cite{golan2015scaling, shetty2013building},
reducing tail latency \cite{balmau2019silk, luo2019performance, sears2012blsm},
improving memory management mechanisms \cite{bortnikov2018accordion, kim2020robust, luo2020breaking},  
and exploiting data characteristics \cite{absalyamov2018lightweight, ren2017slimdb, yang2020leaper}. These works offer promising improvement in various aspects. 

Most of the works~\cite{sears2012blsm, dayanspooky, sarkar2020lethe, alkowaileet2019lsm, dayan2021chucky, zhang2018elasticbf, zhu2021reducing, dayan2017monkey, luo2020rosetta, knorr2022proteus, zhang2018surf, ahmad2015compaction, huang2019x, vinccon2018noftl, zhang2020fpga, wang2014efficient, thonangi2017log, chan2018hashkv, lu2017wisckey, balmau2017triad, golan2015scaling, shetty2013building, wu2015lsmtrie, balmau2019silk, luo2019performance, bortnikov2018accordion, kim2020robust, luo2020breaking, absalyamov2018lightweight, ren2017slimdb, yang2020leaper, alsubaiee2014asterixdb}, unfortunately, are not from the angle of optimizations under dynamic workloads, because their designs are not workload-aware. Recent studies working towards workload-aware LSM-trees lie in tuning the {\it compaction policy}~\cite{dostoevsky2018, idreos2019designcontinuum, chatterjee2021cosine, huynh2021endure, dayan2019log}. 
Compaction improves the read efficiency in an LSM-tree because it reduces the sorted runs to be probed in each query and cuts down space by eliminating redundant entries. However, compaction itself is a costly process that affects write efficiency. The more aggressive in compaction, the lower read amplification and the higher write amplification. 
 Based on these observations, Dayan and Idreos~\cite{dostoevsky2018} further point out that a more frequent compaction optimizes the lookup cost but increases the update cost, and hence, various customized compaction policies based on workload characteristics are proposed to improve the overall performance of the LSM-tree.

There are still several deficiencies in these closest works when applied to dynamic workloads. 
Most of them~\cite{chatterjee2021cosine,dostoevsky2018,idreos2019designcontinuum} compute an optimized system setting when given a specific application workload. 
For example, Cosine~\cite{chatterjee2021cosine} and Dostoevisky~\cite{dostoevsky2018} require users to input a workload with a specific ratio between lookups and updates, so that they can model the expected performance of various possible compaction policies, and choose the best one. While these models are very promising in analyzing a workload-aware system, it remains open regarding how to design reasonable key-value stores when the workload is not known a priori. One possibility is to combine them with workload forecasters such as those designed by Ma {\it et al.}~\cite{ma2018query}, but this incurs an additional forecaster training cost.
{More importantly}, even though the additional workload forecaster may be integrated, the obstacle is in efficient adaptation of the underlying LSM-tree from one state to another in response to workload changes -- the main bottleneck of designing dynamic LSM-trees. For example, changing a compaction policy at a level usually requires resizing the runs and remapping each data block to the runs, which incurs a high cost. This issue has been largely overlooked in the literature and the only known strategy is a greedy transition mentioned by Dayan and Idreos~\cite{dostoevsky2018} as an extended discussion. Nonetheless, this strategy is sub-optimal as we discuss in Section~\ref{sec:flsm_prop}. 

\subsection{The Problem and Our Solution}\label{sec:problem}
We aim to design an LSM-tree based key-value store that can automatically adapt itself in an online manner to continuously changing workloads while keeping a robust end-to-end performance (i.e., close to the performance of the optimal setting in the hindsight).
In contrast to prior works on autonomous DBMS, our focus on NoSQL key-value stores (particularly, LSM-trees) brings in drastically different technical challenges. We thus explore new ways of modeling key-value stores so as to achieve our goal. 
Below we give a preview of our new ideas. 

\vspace{1mm}
\noindent
{\bf Modeling Key-Value Stores Using Reinforcement Learning.} We model (See Section~\ref{sec:overview} and Section~\ref{sec:lerp}) the LSM-tree based key-value store as a reinforcement learning process~\cite{sutton1998introduction, sutton2018reinforcement}, which is a machine learning approach that trains a model with an action-reward series in an online manner. We remark that most existing methods for tuning compaction policy \cite{dostoevsky2018, huynh2021endure, idreos2019designcontinuum, chatterjee2021cosine, dayan2019log} use a different way in modeling, which captures the relationship between compaction policy and LSM-tree performance via a sophisticated formula, referred to as a white-box cost model.
While these models show promising results in many aspects, there are limitations in modeling the end-to-end system performance. { First, }white-box models often focus on I/O complexity analysis to predict the I/Os incurred by each lookup or update, neglecting the CPU overheads that can be significant. For example,
Zhu {\it et al.}~\cite{zhu2021reducing} reported that the hashing of Bloom filters in LSM-trees may incur high CPU overhead that even exceeds I/O overhead with modern storage hardware. Neglecting the CPU overhead is therefore a source of inaccuracy. {Second,} modern systems are extremely sophisticated in design, and are often cumbersome, if not unpromising, to be unfolded for detailed complexity analysis of internal components. For example, memory cache can significantly affect the performance, but white-box formulas are often unable to model such bottom-level details. 

In contrast to white-box formula based models, models based on RL for tuning treat the system as a black-box, where {\it action} modifies the LSM-tree compaction policies and {\it reward} traces system performance. The tuning is automatic in a machine-learning style. Modeling the process with RL is not trivial, as discussed in Section~\ref{sec:lerp}. 

It is also important to note that we do not aim to completely replace classic white-box models which are still helpful in many aspects. Instead,
our main purpose is to explore the potential of applying RL in tuning LSM-tree based systems,
which complements the existing classic methods and, possibly, opens the new research direction for machine-learning-based LSM-tree tuning.

\vspace{1mm}
\noindent
{\bf Designing Transition-Friendly LSM-trees for Reinforcement Learning.} 
When dynamically tuning LSM-trees, one critical issue is maintaining an efficient transformation of LSM-trees, as a slow transformation may outweigh the benefit of a compaction policy adjustment. This issue motivates us to design {\FLSM} (See Section~\ref{sec:flsm}), a flexible LSM-tree that allows various-sized runs in a level. When the compaction policy changes, {\FLSM} avoids the modification of most of the data in the affected level. We also provide a theoretical analysis showing that {\FLSM} promises zero delay, free immediate transition costs, as well as less implicit additional costs. 

\vspace{1mm}
\noindent
{\bf Integrating White-box Models in Reinforcement Learning.}
Unlike RL in other typical machine learning scenarios, it is unpromising to gather a large amount of training samples in a key-value store because the samples are obtained by running costly lookup and update operations. To circumvent the problem, we propose a new {\it level-based} training strategy, where the compaction policies of smaller levels are learned by RL while those of large levels are extended by the lower-level policies through a careful complexity analysis (i.e., a white-box model). This technique largely reduces the number of samples needed to train the model, empowering the tuning ability of RL.

Putting together, we develop the {\ourmethod} system, a flexible LSM-tree-based key-value storage system that is self-adaptive to dynamic workloads. 
We summarize our contributions as follows. 

\begin{itemize}[leftmargin=*]
    \item  We make the first attempt at designing RL models to optimize the performance of LSM-tree based key-value systems. We design and implement {\ourmethod} which is a key-value system that demonstrates strong robustness in end-to-end performance via an RL-based tuning model. {\ourmethod} can efficiently adapt itself to various workloads, based on the instructions of the learning model.
    \vspace{2mm}
    \item  We address an open problem in dynamic key-value systems -- the efficient and transient policy transformation in LSM-tree under dynamic workloads. We present {\FLSM{}}, a novel LSM-tree structure that incorporates a flexible compaction policy transition scheme, which swiftly transforms the compaction policy of an LSM-tree with minor additional overhead.
    \vspace{2mm}
    \item  We present a new level-based RL model which is embedded with a policy propagation formula to reduce the number of samples required, shrink the action space and improve the convergence speed in our {\ourmethod} system. 
    \vspace{2mm}
    \item We conduct extensive experimental evaluations of {\ourmethod} in real system environments. 
    We demonstrate that {\ourmethod} achieves high performance in all the tested workloads. {\ourmethod} achieves up to 4x throughput compared with the baselines.
\end{itemize}

\vspace{-2mm}

\section{Background}
{
\renewcommand\arraystretch{1.15}
\addtolength{\tabcolsep}{1.4pt}
\begin{table}[b]
\vspace{4mm}
  \addtolength{\tabcolsep}{-1pt}
  \scalebox{0.95}{
  \begin{tabular}{ll}
  \hline
  \multicolumn{1}{|l|}{\textbf{Notation}}  & \multicolumn{1}{l|}{\textbf{Description}}                                 \\ \hline

  \multicolumn{1}{|l|}{$T$}              & \multicolumn{1}{l|}{Capacity ratio between adjacent levels}                     \\ \hline
  
  \multicolumn{1}{|l|}{$K_i$}                & \multicolumn{1}{l|}{Compaction policy of the $i$-th level}           \\ \hline

  \multicolumn{1}{|l|}{$E$}          & \multicolumn{1}{l|}{Size of a key-value entry (in bytes)}             \\ \hline

  \multicolumn{1}{|l|}{$B$}          & \multicolumn{1}{l|}{Size of a disk page (in bytes)}             \\ \hline

   \multicolumn{1}{|l|}{$C_i$}          & \multicolumn{1}{l|}{Capacity of the $i$-th level (in bytes)}             \\ \hline

   \multicolumn{1}{|l|}{$f_i$}          & \multicolumn{1}{l|}{False positive rate of Bloom filters at the $i$-th level}             \\ 
   \hline

   \multicolumn{1}{|l|}{$D_i$}          & \multicolumn{1}{l|}{Total data size in the $i$-th level (in bytes)}             \\ 
   \hline
                                                            
  \end{tabular}
  }
  \vspace{1mm}
  \caption{Frequently used notations. }\label{table:notation}
  \vspace{-5mm}
  \end{table}
  }
  \setlength{\textfloatsep}{0pt}

\vspace{1mm}
\noindent\textbf{LSM-tree.} The log-structured merge tree (LSM-tree)~\cite{o1996log} is a persistent multi-level index structure that stores key-value entries. When a key-value entry is inserted, an LSM-tree buffers it in the main memory. Once the buffer is filled up, it sorts the entries by keys and flushes them into a disk-resident sorted run in Level 1. When the size of the disk runs in a level exceeds a certain threshold, the LSM-tree sort-merges them into a larger run and puts it into the next level. This process of sort-merging multiple smaller sorted runs in Level $i$ to a larger sorted run in Level $i+1$ is called {\it compaction}. As such, the capacity of Level $i+1$ is $T$ times larger than that of Level $i$, where $T$ is the capacity ratio. 

\vspace{1mm}
\noindent\textbf{Lookup (Query) in LSM-tree.} 
In modern key-value systems, LSM-tree is coupled with Bloom filters and fence pointers to improve the key lookup performance. The Bloom filter and fence pointer are built for each sorted run and will be loaded into the main memory when the system starts.
To check whether a key exists in a run, the Bloom filter associated with the run is first probed. If it returns false, then the run does not contain the key and no I/O cost will be incurred. Otherwise, accessing the run in the disk is necessary, which incurs $O(1)$ I/Os guaranteed by fence pointers. 
In general, the Bloom filter largely prevents unnecessary disk reads.
The I/O-cost reduced by the Bloom filter depends on its FPR (False-Positive Rate). In particular, if the run does not contain the queried key, its Bloom filter may still mistakenly return true for the key, incurring a false positive. 
While mainstream designs uniformly set the FPRs for Bloom filters at all LSM-tree levels, state-of-the-art techniques \cite{dayan2017monkey,dostoevsky2018,huynh2021endure,dayan2019log,chatterjee2021cosine} employ the {\Monkey} allocation scheme \cite{dayan2017monkey}, which arranges exponentially higher FPRs to
Bloom filters in larger levels.

\vspace{1mm}
\noindent\textbf{Compaction Policy.} 
The compaction policy $K_i$ at each Level $i$ can be independent, as presented in Dostoevsky~\cite{dostoevsky2018}. The policy $K_i$ means that there are at most $K_i$ sorted runs in the level.
Level $i$ has an active run that admits the merged run from Level $i-1$.
The capacity for each active run is $1/K_i$ of the level capacity.
All entries in a level are eventually merged and flushed down to the next level when the level reaches its capacity. 
Dostoevsky proposes {\lazyl} which applies an aggressive compaction policy of at most 1 sorted run in the largest level while all other levels use a lazy compaction policy in common.
They give a way of applying different compaction policies at different levels, but within the same level the compaction policy is always the same. Our framework give a significant extension by allowing various run sizes in a level. 

\vspace{1mm}
\noindent\textbf{Amplification.} 
Read amplification and write amplification are commonly used to measure the performance of an LSM-tree. Read amplification refers to the average number of disk pages read during a lookup, since a query may probe multiple sorted runs in an LSM-tree to search for the queried key. Write amplification refers to the average number of physical rewrites that an entry incurs, because each data entry is constantly rewritten to storage as a result of the LSM-tree's compaction. 

\section{{\ourmethod} Overview}\label{sec:overview}

This section presents {\ourmethod}, a key-value system empowered with deep reinforcement learning. 
The purpose of {\ourmethod} is to tune the compaction policy in each LSM-tree level to best process the workload. There is no doubt that a fixed policy is still possible to be applied under dynamic workloads, but as we have mentioned, changing policies can bring in benefits as long as the LSM-tree structure transformation cost is small. {\ourmethod} includes two major designs: (1) A level-based deep reinforcement learning tuning model, named {\bf {\RLmodel}} and (2) A new flexible LSM-tree with new compaction transition method, named {\bf \FLSM{}}. As we will explain shortly, {\RLmodel} incorporates the cost analysis across levels for fast convergence and less sample-demanding training. {\FLSM} embeds flexible transition to avoid high transition cost, hence being more cost-friendly to a compaction policy tuner like {\RLmodel}. We will first 
describe the general workload of {\ourmethod} in this section, and then discuss the detailed mechanism of {\FLSM} and {\RLmodel} in Section~\ref{sec:flsm} and Section~\ref{sec:lerp}. Table~\ref{table:notation} lists the frequently used notations.

\begin{figure}[h]
\vspace{-1mm}
    \centering
    \includegraphics[width=0.75\textwidth]{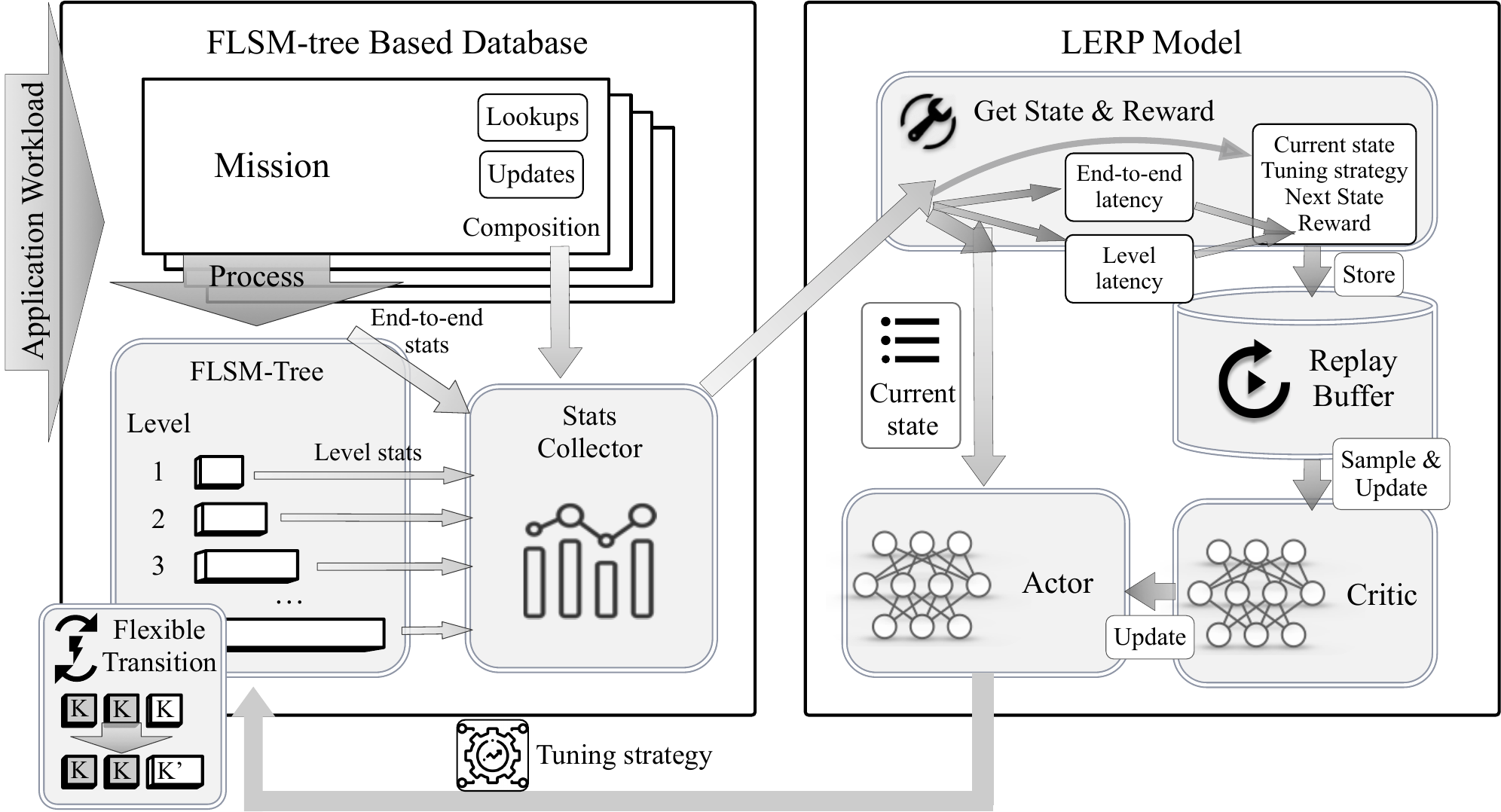}
    \vspace{-2mm}
    \caption{The components and workflow of {\ourmethod}.}
    \vspace{-4mm}
    \label{fig:basic intro}
\end{figure}
\begin{figure*}[t]
    \centering
    \vspace{0mm}
    \includegraphics[width=1.00\textwidth]{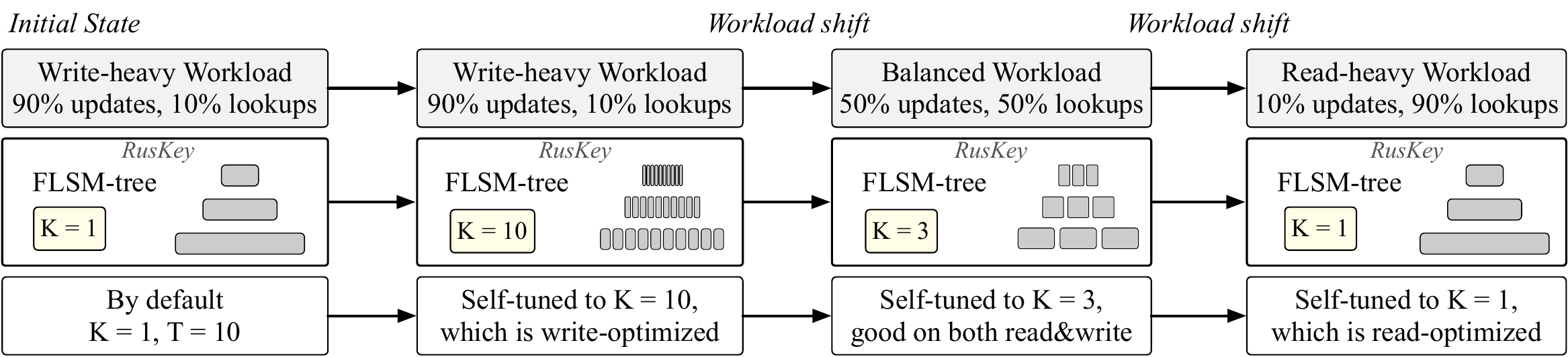}
    \vspace{-7mm}
    \caption{\color{black} Running example of {\ourmethod}. This is a simplified example where we assume that each level has the same compaction policy. In fact, {\ourmethod} allows the {\FLSM} to set various compaction policies across different levels.}
    \label{fig:ruskey running example}
    \vspace{1mm}
\end{figure*}

\subsection{System Workflow}
{\color{black}
In {\ourmethod}, the \textit{application workload} is a series of {\it lookups} or {\it entry updates} operations input to the database. A lookup finds the corresponding value for a given key, and an update inserts or deletes a key-value entry. {\ourmethod} allows for arbitrary arrivals of operations, which means that knowledge of the query/update ratio is not known in advance. 
A {\it mission} in {\ourmethod} is a collection of operations (e.g., lookups or updates) in the workload. Conceptually, the workload is divided into missions. After processing each mission, {\ourmethod} may adjust its internal LSM-tree structure for the coming missions. {\ourmethod} also maintains a {\it statistics collector} that keeps track of necessary statistics of {\ourmethod} and application workload over time. Besides overall statistics of the {\FLSM}, it tracks statistics separately for each {\FLSM} level to support the level-based training scheme in {\RLmodel}. 
It also collects the operation composition in each mission for detecting changes in the application workload.

Figure~\ref{fig:basic intro} illustrates the workflow of {\ourmethod}.
Suppose a real-world application employs {\ourmethod} as its data storage backend. {\ourmethod} processes operations from the application workload with the {\FLSM{}} and uses the tuning model {\RLmodel} to adjust the compaction policy. {\RLmodel} and {\FLSM{}} interact after each mission. At the end of a mission, {\RLmodel} gives tuning suggestions to the {\FLSM} and updates the {\network}. 
Meanwhile, the statistics collector sends statistics of {\FLSM} in this mission to \RLmodel, 
which extracts experience samples from the statistics and stores them in the replay buffer. The replay buffer stores some quadruples (called experience samples), each of which contains the essential information for reinforcement training. The information includes the action of policy change, the {\FLSM} state before the action, the {\FLSM} state after the action, and the subsequent mission, as well as the reward of the action. Through learning from these samples, {\RLmodel} effectively updates its internal {\network}~\cite{lillicrap2015continuous}. Then {\RLmodel} selects a suitable tuning strategy for the compaction policy of the {\FLSM} in the next mission according to the current state of the {\FLSM}.
{\ourmethod} then updates the {\FLSM} based on the suggested compaction policy with the flexible transition. 
Under a stable application workload, the {\network} converges in a number of missions, shifting {\FLSM{}} towards the optimal compaction policy. Once the application workload changes, the {\network} is no longer in convergence, and {\RLmodel} will restart to exploit compaction policies under the new workload.}

\vspace{1mm}
\noindent{\bf Running Example.}
Figure~\ref{fig:ruskey running example} shows a running example of {\ourmethod}, which is simplified by assuming each level has the same compaction policy. 
Initially, the compaction policy is set as 1, and the capacity ratio is 10. The dynamic workload is write-heavy at first with 90\% of updates and 10\% of lookups. It then shifts into a read-write-balanced workload (50\% of updates and 50\% of lookups), followed by a read-heavy workload (10\% of updates and 90\% of lookups). At the beginning, {\ourmethod} tunes the compaction policy of the {\FLSM} from $K=1$ to $K=10$ to achieve the best write performance under write-heavy workloads, and we note that during the tuning the system is not fed with the workload statistics. With the benefit of flexible transition, the {\FLSM} could transform its policy $K=1$ to $K=10$ without a transition cost and delay. Once the workload shifts to the balanced one, {\ourmethod} will re-tune the compaction policy of the {\FLSM} from $K=10$ to $K=3$, which is more effective in a balanced workload. Similarly, when the workload shifts to read-heavy, {\ourmethod} will adjust its compaction policy from $K=3$ to $K=1$ to achieve an optimized read efficiency.

\vspace{-2mm}
\section{{\FLSM} in {\ourmethod}}\label{sec:flsm}

This section discusses two baseline transition methods (Section~\ref{sec:base_transit}), followed by presenting our design of {\FLSM} (Section~\ref{sec:main_design_flsm}) and its desired properties (Section~\ref{sec:flsm_prop}).

\subsection{Greedy and Lazy Transitions}~\label{sec:base_transit}
{\ourmethod} changes the underlying LSM-tree based on the suggested compaction policy by the tuning model. However, in the classic LSM-tree, continuously changing the compaction policy demands the reallocation of memory and potentially reorganizes the data in the LSM-tree. Enforcing a change of compaction policy at a level can have two straightforward solutions, namely, {\it greedy transition} and {\it lazy transition}. However, these solutions are not ideal.

\begin{figure}[b]
\vspace{1mm}
    \centering
    \hspace{4mm}
    \makebox[0pt][c]{
    \includegraphics[width=0.66\textwidth]{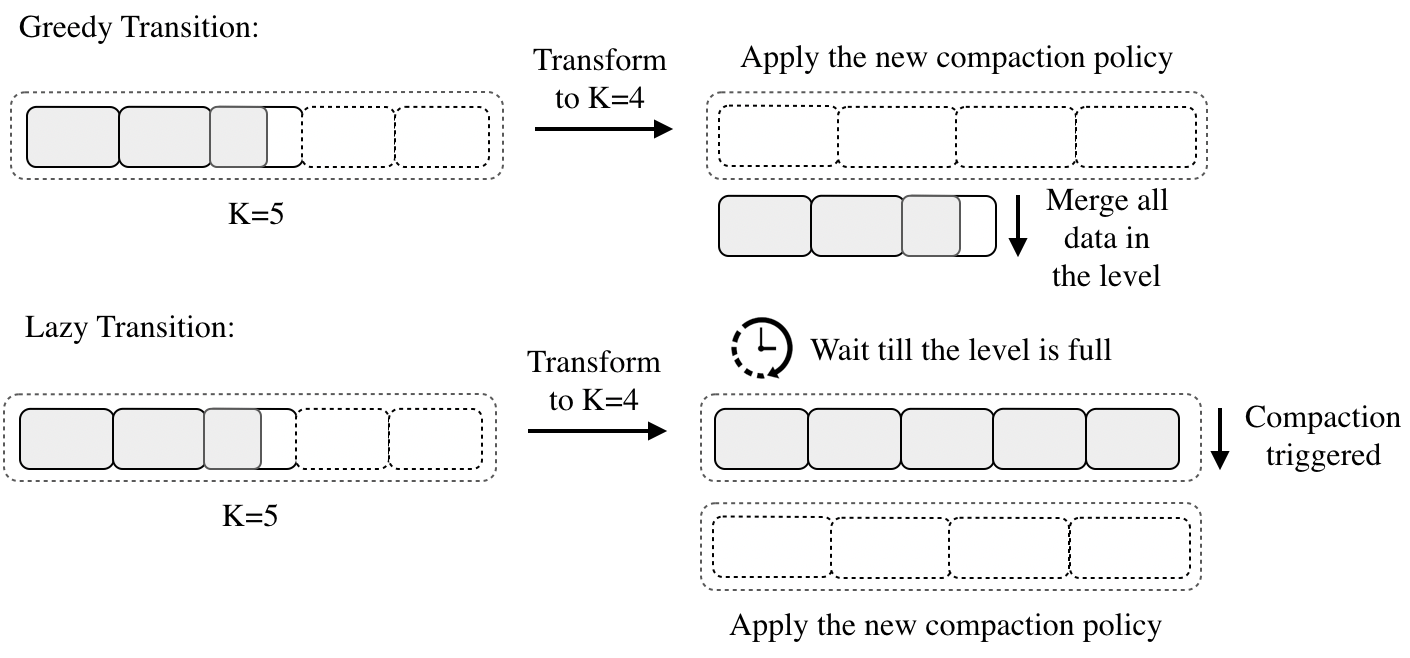}
    }
    \vspace{-2mm}
    \caption{Greedy and lazy transitions of compaction policy.}
    \label{fig:basic transitions}
\end{figure}

\vspace{1mm}
\noindent {\bf{Greedy Transition}~\cite{dostoevsky2018}.}
Once a level receives a tuning strategy, it merges and flushes all its data into its next level. The current level then needs to be rebuilt based on the new compaction policy, as shown in Figure~\ref{fig:basic transitions} (top). We note that such a greedy transition is not effective due to its dreadful cost and the write stall incurred during data flush, and hence significantly degrades the performance. Moreover, if an unfilled level is merged with the next level, it incurs a larger write amplification than an LSM-tree without transitions. 

\vspace{1mm}
\noindent {\bf{Lazy Transition.}}
The other extreme is the lazy transition that an LSM-tree level changes its compaction policy only after it is merged to the next level, as shown in Figure~\ref{fig:basic transitions} (bottom). Although such a lazy transition will not bring an immediate compaction cost, it affects the timeliness of the information fed into the RL model. To explain, a given (policy change) action may have no effect on the LSM-tree level until the level eventually becomes full and gets merged, and the delay can be long at a large level, ultimately reducing the effectiveness of the model.

\begin{figure*}[t]
\vspace{0mm}
    \centering
    \makebox[0pt][c]{
    \includegraphics[width=0.99\textwidth]{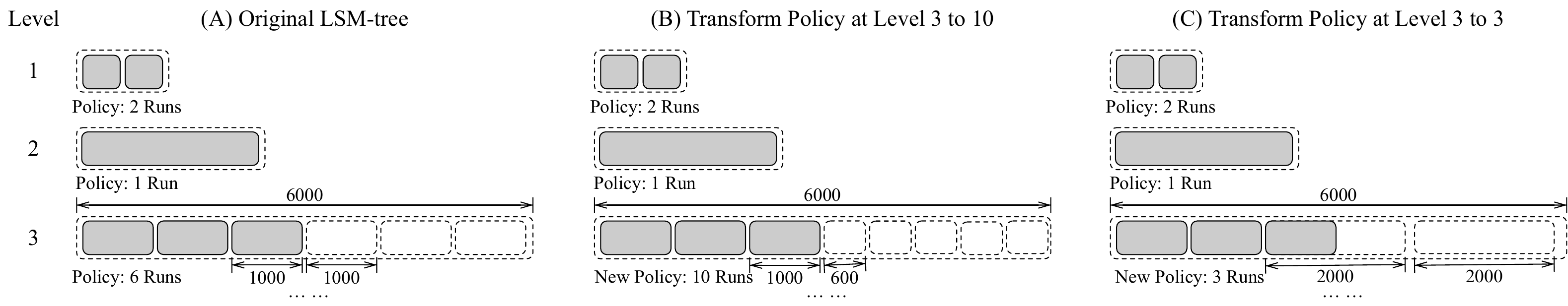}
    }
\vspace{-4.3mm}
    \caption{Flexible compaction policy transition in a FLSM-tree.}
    \label{fig:flexible transition}
\vspace{-5mm}
\end{figure*}

\subsection{Main Design of {\FLSM}}\label{sec:main_design_flsm}
We design a {\bf{F}}lexible {\bf LSM}-tree, abbreviated as {\FLSM}, to accommodate a flexible transition between compaction policies.  
A novel design in {\FLSM} is that it allows different-sized runs to exist at the same level. This extends the classic LSM-tree structure, allowing more room for an LSM-tree to morph the structure gradually. At each level, there is an active run that can admit more key-value entries. When the active run in the level reaches its capacity, it is sealed, and a new active run is created. When a new policy is applied on a level, it never affects the runs that have been sealed, and only changes the capacity of the active run and the runs to be formed later. In this way, the change of the policy is immediate and the {\RLmodel} model gets the correct reward by applying the policy changes without incurring any immediate transition costs. 

Let the capacity of Level $i$ be $C_i$, the new compaction policy be $K'_i$, 
and the original compaction policy of Level $i$ is $K_i$. 
The discussion is divided into two cases.

\vspace{1mm}
\noindent
{\bf If \bm{$K_i^{'}<K_i$}}, {\FLSM} will increase the capacity of the active run, keeping the other runs intact.

\vspace{1mm}
\noindent
{\bf If \bm{$K_i^{'}>K_i$}}, {\FLSM} will decrease the capacity of the active run. One may have a concern that when the new policy is known, the active run size can already be larger than the expected size $C_i/K_i^{'}$. In this case, we immediately seal the active run and create a new active run. The runs, when sealed, remain untouched until the level holding them reaches its capacity and triggers a compaction during which all runs will be merged to the next level. 

\vspace{1mm}
\noindent
{\bf Example.}
{
Figure~\ref{fig:flexible transition} presents examples of transforming compaction policy in {\FLSM}. 
A grey square represents a sorted run filled with data, and the dashed line denotes logical capacity of a level or an incoming run. In Figure~\ref{fig:flexible transition}~(A), the compaction policy of Level 1 is 2. Level 1 includes two runs, each having a size that is half of the level capacity. The compaction policy of Level 2 is 1, rendering a single run in the level. For Level 3, initially the compaction policy is 6 and the level capacity is 6000, giving an expected run size 6000/6=1000. 
If we transform the policy of Level 3 to 10, it will turn to the state of Figure~\ref{fig:flexible transition}~(B), such that the expected run size under the new compaction policy would not affect existing sealed runs, but the expected size of an incoming run decreases to 600 (=6000/10).
If we transform the policy to 3, it will turn to the state of Figure~\ref{fig:flexible transition}~(C). The expected size of an incoming run becomes 6000/3=2000, and the current active run size also increases to $2000$.
}

\begin{figure*}
\vspace{-2mm}
  \makebox[0pt][c]{
    \hspace{0mm}
    \begin{minipage}[h]{0.5\textwidth}
    \vspace{3mm}
            \centering
\renewcommand\arraystretch{1.5}
   \addtolength{\tabcolsep}{-2.4pt}
 \scalebox{0.7}{
 \begin{tabular}{|ll|c|c|c|}
\hline
\multicolumn{2}{|l|}{\textbf{Transition Alg.}}  & \multicolumn{1}{c|}{Greedy } & \multicolumn{1}{c|}{Lazy } & \multicolumn{1}{c|}{\FLSM\ (ours)} \\ \hline
\multicolumn{2}{|l|}{\textbf{Transition Cost}}              & 
  \multicolumn{1}{c|}{$\frac{C}{2B}$}     &
   \multicolumn{1}{c|}{$0$}  &
   \multicolumn{1}{c|}{$0$}    \\ \hline
\multicolumn{2}{|l|}{\textbf{Delay (secs)}}              & 
   \multicolumn{1}{c|}{$0$}   &     
   \multicolumn{1}{c|}{$\frac{C}{2N_u E}$}    &
   \multicolumn{1}{c|}{$0$}   \\   \hline
\multicolumn{1}{|l|}{\multirow{2}{*}{\begin{tabular}[c]{@{}l@{}}\textbf{Additional}\\\textbf{Cost}\end{tabular}}} & 
$K<K'$  &
\multicolumn{1}{c|}{$\frac{TC(1-x)}{2BK}$}     &
\multicolumn{1}{c|}{$\frac{TC(1-x)(K'-K)}{2BKK'}$}  &
\multicolumn{1}{c|}{$0$}  
\\ \cline{2-5} 
\multicolumn{1}{|l|}{}     &
\multirow{2}{*}{$K>K'$}& 
\multicolumn{1}{c|}{$\frac{TC(1-x)}{2BK}$}     &
\multicolumn{1}{c|}{$\frac{fC(1-x^2)(K-K')\gamma}{2E(1-\gamma)}$}  &
\multicolumn{1}{c|}{$\frac{fC(x-x^2)(K-K')\gamma}{E(1-\gamma)}$}   
 \\ \cline{1-1} \cline{3-5} 
 \multicolumn{1}{|l|}{\textbf{Case Study}}  &            & 
   \multicolumn{1}{c|}{$125$ I/Os}   &     
   \multicolumn{1}{c|}{$3.75$ I/Os}    &
   \multicolumn{1}{c|}{$2.5$ I/Os}   \\   \hline
\end{tabular}
}
  \vspace{2mm}
  \captionof{table}{Transition costs and delays. 
  }\label{table:complexity}
        \end{minipage}
        \hspace{0mm}
        \begin{minipage}[h]{0.5\textwidth}
            \centering
    \vspace{5mm}
    \makebox[0pt][c]{
    \includegraphics[width=0.9\textwidth]{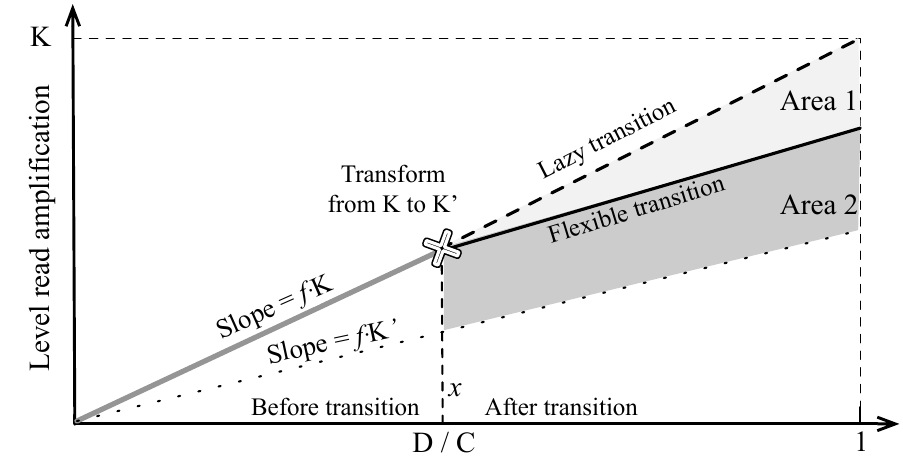}
    }
    \vspace{-3mm}
    \caption{Flexible transition incurs an additional cost = Area 2; lazy transition incurs an additional cost = Area 1 + Area 2.}
    \label{fig:implicit}    
        \end{minipage}
}
\vspace{2mm}
\end{figure*}

\subsection{Desired Properties of {\FLSM}}
\label{sec:flsm_prop}
To check the advantages of our {\FLSM} design, we focus on the following measures: (1) {\it Transition Cost}, which is the immediate compaction cost incurred by a transition. We aim at small immediate transition cost to avoid serious performance glitches; (2) {\it Delay}, which is the time span between the new policy's arriving at the system and the
time at which the transition actually takes effect on the LSM-tree; (3) {\it Additional Cost}, which is the total additional I/O cost incurred by a transition, as compared to an ideal LSM-tree born with the new compaction policy.

{
We illustrate the concept of {\it additional cost} incurred by transition using Figure~\ref{fig:flexible transition} (C) as an example. After transforming the policy in Level 3 from 6 to 3, the incoming runs will follow the new compaction policy 3, while the existing sealed runs stay with the previous compaction policy 6. In this instance, there will be at most 4 runs in Level 3 after the flexible transition, whereas the new compaction policy indicates 3 runs in the level, such that accessing the level would incur more I/Os than expected. {\color{black}Here we only discuss the additional cost, and hence the cost is lower bounded by $0$. We note that sometimes additional benefits may be involved, for which we will have a further discussion in the {\it overall comparison} paragraph at the end of this section.}
}

\vspace{1mm}
\noindent
{\bf Notations.} To facilitate our discussion, we first introduce the necessary notations for the analysis in this sub-section. We fix an LSM-tree level whose capacity (in bytes) is $C$. 
Let $D$ be the total data amount (in bytes) of key-value entries currently stored in the level. Let $B$ be the size (in bytes) of a disk page. Let $E$ be the size (in bytes) of a key-value entry. Let $N_u$ be the number of updates arriving per second. Let $x$ be the percentage of the stored data in the current level when a compaction policy transition is applied. Let $f$ be the FPR of the Bloom filter in the level. Let $\gamma$ denote the percentage of lookups in the workload.

\vspace{1mm}
\noindent\textbf{{\FLSM} incurs zero transition cost and delay.}
Our {\FLSM} has the minimum transition cost and delay cost. 
The greedy transition merges all data in that level to the next level. This process incurs $\frac{D}{B}$ I/Os.
When a transition occurs, $D$ may be any random value between $0$ and $C$, leading to an amortized I/O cost $\frac{C}{2B}$ for a greedy transition. 
In contrast, the transition cost of a lazy transition is always zero. As for the delay, in the lazy transition, the new compaction policy takes effect only after the level is full and a whole-level compaction is triggered. The delay can be estimated by $\frac{C-D}{N_u E}$ seconds. 
The amortized delay of the lazy transition is $\frac{C}{2 N_u E}$ considering $D$ ranges from $0$ to $C$.
Regarding the transition mechanism in a \FLSM, the transition cost is always zero, since it only changes the metadata of the active run and does not incur any disk reads. 
{\color{black} Meanwhile, the flexible transition in a {\FLSM} comes into effect immediately with zero delay}{\footnote{\color{black} Zero-delay means the LSM-tree immediately reacts when policy changes, marking that upcoming updates will follow the new policy.  
We note that the policy changes will not have any effect for the extreme case where the workload only contains reads.
In this case, flexible transition simply degenerates into lazy transition and will not bring additional negative impact.}. 
}
Table~\ref{table:complexity} summarizes the transition cost and delay of the three transition methods.

\vspace{1mm}
\noindent\textbf{{\FLSM} incurs less additional cost.} 
{
FLSM-tree also incurs the minimum additional costs compared with baselines. We first examine the additional cost of each transition method.

\vspace{1mm}
\noindent
{\it \underline{Additional cost of the greedy transition.}}
A greedy transition compacts the level right after the policy transition, giving zero additional read cost. However, this compaction suffers a larger write amplification because it merges fewer data into the next level than a normal full-level compaction. 

We now analyze the additional write amplification incurred by a greedy transition. 
Let us first consider a normal compaction without policy changes. Without loss of generality, we assume that the subsequent level's compaction policy is $K$. In expectation, the size of the active run in the subsequent level is $\frac{T C}{2K}$. 
A compaction reads expected $\frac{T C}{2K}$ bytes of data and rewrites them into the disk in order to merge $C$ bytes of data (i.e., the full current level) into the subsequent level.
Hence, the active run size $\frac{T C}{2K}$ divided by $C$, which is $\frac{T}{2K}$, is the write amplification of the current level.   
The additional cost caused by the greedy transition is $\frac{T}{2K\cdot x}$ because it only merges data that is $x$ times the level capacity, which leads to $\frac{T}{2K\cdot x}-\frac{T}{2K}$ additional write amplifications. 

The expected additional I/O cost is computed by multiplying the additional write amplification with the number of pages merged by the greedy transition into the subsequent level:
\begin{equation}
    Cost_{greedy}=\left(\frac{T}{2K\cdot x}-\frac{T}{2K}\right)\cdot\frac{C\cdot x}{B}= \frac{TC(1-x)}{2BK}
\label{eq: greedy cost}
\end{equation}

\vspace{1mm}
\noindent
{\it \underline{Additional cost of lazy transition.}}
If $K>K'$, lazy transition avoids the immediate transition cost and the additional cost from write amplification, at the expense of incurring additional read cost. 

We illustrate this expense in Figure~\ref{fig:implicit},
where the horizontal-axis refers to the level's filling ratio, which is the percentage of data stored at the level among the capacity of the level, i.e., $D/C$. 
The vertical-axis refers to the read amplification of the level, which is the expected number of disk pages a lookup needs to probe in the level.
In a process of inserting data from an empty level to a full level, the filling ratio increases from $0$ to $1$. 
In expectation, the read amplification of a level is the product of three values -- its false positive rate $f$, its compaction policy $K$, and its filling ratio. To explain, when the level is full, there are $K$ runs in the level to be probed during each query and each run probe requires a disk I/O only when its Bloom filter returns positive. Therefore, the read amplification can be expressed by $f\cdot K\cdot\frac{D}{C}$. 

As shown in Figure~\ref{fig:implicit}, after the transition, the read amplification of the level is higher than it should be under the more aggressive compaction policy $K'$ (which is represented by the dotted line). 
This is because the level still adheres to the foregoing $K$ compaction policy with lazy transition.
This lasts until the level is full, at which point all data is merged into the following level via compaction.
Therefore, the sum of Area 1 and Area 2 in Figure~\ref{fig:implicit} represents the overall additional read cost of a lazy transition. 
A lazy transition's additional cost in I/Os can be expressed by Equation~\ref{eq: lazy cost}. 
\begin{equation}
    Cost_{lazy}=\frac{fC(1-x^2)(K-K')\gamma}{2E(1-\gamma)}
\label{eq: lazy cost}
\end{equation}

If $K<K'$, a lazy transition incurs an additional write cost.
Upon changing the policy, the expected write amplification for this level becomes $\frac{T}{2K'}$. However, until the level becomes full and triggers compaction, the actual write amplification of the lazy transition remains $\frac{T}{2K}$. 
Throughout this period, this level will merge a total of $\frac{(1-x)C}{B}$ blocks, incurring an additional cost of $\frac{TC(1-x)(K'-K)}{2BKK'}$. 

\vspace{1mm}
\noindent
{\it \underline{Additional cost of a flexible transition.}}
The additional cost of a flexible transition can be derived in a similar way to a lazy transition.
Before changing a compaction policy, the growth rate of read amplification equals the current compaction policy $K$. When a new compaction policy $K'$ is applied, different transition methods bring in distinctions. 
With a lazy transition, the growth rate of read amplification stays the same after the change of the policy.
With a flexible transition, the growth rate reduces from $f\cdot K$ to $f\cdot K'$.
Similar to a lazy transition, a flexible transition also incurs additional read amplification when the level is full.
This is because part of the data in this level is inserted before the transition, where the inserted data adheres to the foregoing compaction policy $K$.
Therefore, Area 2 in Figure~\ref{fig:implicit} represents the overall additional cost of flexible transition, which is 
\begin{equation}
    Cost_{flexible}=\frac{fC(1-x)(K-K')\gamma}{E(1-\gamma)}
\label{eq: flexible cost}
\end{equation}

\noindent \underline{\it Overall Comparison.} 
Table 2 summarizes the additional cost of different transition methods.
{It is obvious that the additional cost of lazy transition is always larger than flexible transition. 
} For comparison between flexible transition and greedy transition, it is obvious that greedy transition is always more costly when $K<K'$ because the additional cost of a flexible transition is zero in this case. When $K>K'$, 
by removing the constant coefficients in cost functions (Equations~\ref{eq: lazy cost},~\ref{eq: flexible cost}), the additional costs of a flexible transition and a greedy transition are $O(\frac{fC}{E})$ and $O(\frac{C}{B})$, respectively. 
Note that the FPR $f$ in an LSM-tree is often less than $1\%$, and the entry size $E$ is usually in the same magnitude as the disk page size $B$. Therefore, the cost of a flexible transition can be up to two orders of magnitude smaller than that of a greedy transition. 

We give an example in real scenarios for ease of comparison. We let $x=1/2$ and $\gamma=1/2$ since $x$ and $\gamma$ are both distributed between $0$ and $1$. We follow the common settings in key-value stores and set $T=10$, $B=4096$, $E=1024$, $C=1024000$, and $f=0.01$. Finally, setting $K=5$ and $K'=4$ gives us $\frac{TC(1-x)}{2BK}=125$, $\frac{fC(1-x^2)(K-K')\gamma}{2E(1-\gamma)}=3.75$, and $\frac{fC(x-x^2)(K-K')\gamma}{E(1-\gamma)}=2.5$. Hence, a flexible transition incurs a much smaller additional cost under this scenario.

{\color{black}Furthermore, if we consider the overall benefit that takes into account both additional costs and additional benefits, flexible transition in general still has a greater benefit than that of lazy transition, assuming that a newly learned policy outperforms the previous one in terms of current workload and has lower latency. Since flexible transition does not involve any delay compared to lazy transition, it allows for a prompt transformation to the new and improved policy, leading to earlier benefits.}
 
\section{\RLmodel{}: The tuning model in {\ourmethod}}\label{sec:lerp}

Reinforcement learning (RL) is a machine learning paradigm that trains an agent to cognize and decipher its environment. The agent gets reward or penalty for each action it takes, and decides the next action based on previous experience. 
The goal of RL is to improve an agent's performance by learning the optimal policy that maximizes the expected cumulative reward. 

It is not a trivial task to design a suitable RL model for tuning an LSM-tree-based key-value system. First, it remains unexplored how to relate the key-value system properties such as LSM-tree compaction policy, the running time cost, the level properties, and the workload characteristics to the fundamental RL elements such as states, actions, and rewards. The model should also avoid over-amplifying the action space and state space. Second, to make {\ourmethod} work in an online manner under dynamic workloads, the model should converge fast. Unfortunately, the high data volume in key-value systems makes the deeper LSM-tree levels experience much longer time to be affected. As a result, the corresponding feedback to the RL model is significantly delayed. Considering that RL requires sufficient feedback from the actual environment to adjust its policy, efficiently and effectively learning the policy for larger levels is notorious challenging. To address the challenge, we design a novel {\bf Le}vel-based {\bf R}einforcement learning model with policy {\bf P}ropagation, abbreviated as \RLmodel. Next, we introduce the level-based model of \RLmodel\ in Section~\ref{sec:per-level}, and policy propagation techniques in Section~\ref{sec:propagation}. 

\subsection{The Level-based Model}\label{sec:per-level}

We model the {\it reward}, {\it state}, {\it environment}, and {\it action} to capture the system process, and then apply a RL procedure. 

\subsubsection{State and Environment}
The {\it state} captures the parameters related to the {\FLSM} and the workload within a mission. Our model state consists of internal statistics of the LSM-tree, such as the number of read and write I/Os, the level capacities, and the current compaction policies at each level. It also includes workload statistics such as the read/write ratio in the previous mission. 
These statistics effectively capture the system state which is related to the system performance.
We model the {\it environment} by the compaction policy 
of a level. For Level $i$, its compaction policy $K_i$ is defined as the maximum number of sorted runs in the level, and obviously $K_i$ needs to be in $[1, T]$. A smaller $K_i$ means a more aggressive compaction policy. Particularly, $K_i=1$ implies a leveling compaction policy, and $K_i=T$ implies a tiering compaction policy. 

\vspace{1mm}

\subsubsection{Action Space: From $O(2^L)$ to $O(L)$} \label{sec:action}
The {\it Action} changes the compaction policy of an {\FLSM{}}. A compaction policy setting of an $L$-level {\FLSM} is a list $(K_1, K_2, \ldots, K_L)$, where $K_i$ refers to the policy of Level $i$. 
\textcolor{black}{
Intuitively, an action can change a policy setting $\textbf{V}=(K_1,K_2,...,K_L)$ to another $\textbf{V}'=(K_1',K_2',...,K_L')$, 
giving a massive action space of size $O(T^L)$, where $T$ indicates the number of policies in a level.} To reduce the action space, our main idea is two-fold. First, the compaction policy is independent among levels, and hence the training can be done separately for each level. 
Second, the characteristics of real-world workloads often change {\it gradually} over time. This hints that the policy should also be roughly {\it continuous}, and a drastic change of the policy may not be preferred. 

Therefore, we design a {\it level-based} model that trains an independent RL model for each level, where the reward, state, and action are only tailored to the corresponding level. Such level-decoupled models have drastically smaller action and state space, and hence easier to train. To explain, the action space in the model at each level (or level model) is only $O(T)$, and the total action space is $O(TL)$. Furthermore, we only allow continuous change of the policy in level models, i.e., the next action for policy $K_i$ can only be one of $K_i - 1$, $K_i$, or $K_i+1$. This further reduces the action space to $O(L)$.

\subsubsection{Reward.} 
In RL, the reward function provides feedback from the environment indicating the correctness of the actions that have been taken.
In {\ourmethod}, the reward function is obliged to reflect the performance of an {\FLSM} level. Our reward function combines both the end-to-end system latency and a specific level-based latency. The former reflects the global impact while the latter reflects the local impact. Let $t_i$ denote the level-based latency of the $i$-th level and $t'$ denote the end-to-end latency of {\FLSM}. {\RLmodel}'s reward function on Level $i$ is 
$\alpha\cdot t_i+(1-\alpha)\cdot t'$. Where $\alpha$ is a parameter between 0 and 1, which controls the weight of $t_i$ and $t'$.
We do not deliberately set different weights to read latency and write latency, because the relative speed of reads and writes is reflected in the reward (end-to-end latency), and automatically learnt in RL.

\subsubsection{Incorporating Actor-Critic Network}\label{sec:ddpg}

With the fundamental mapping between the major RL elements and {\FLSM}, we now discuss the specific design of the RL model. Note that our level-based design can be applied with different RL models, and we select the Deep Deterministic Policy Gradient (DDPG) model~\cite{lillicrap2015continuous}, which has been shown to be more effective compared with the classic models such as DQN~\cite{mnih2013playing}.

{DDPG integrates DQN with the actor-critic network from DPG (Deterministic Policy Gradient). The actor-critic network can be regarded as a temporal difference variation of the policy gradient.}
It parameterizes a critic function and an actor function with neural networks. The critic function maps the current state and action taken into a Q-Value, which is an estimation of the expected reward. The actor function maps the state into a probabilistic distribution of available actions along the direction suggested by the critic function. The policy gradient technique is the foundation for the actor's learning. Comparatively, critics assess the actor's performance by computing the value function. We refer interested readers to the original paper~\cite{lillicrap2015continuous} for more details of the model. 

Under the {\RLmodel} model,
actors are in charge of selecting the best compaction strategy, and critics assist in assessing the effectiveness of a compaction strategy.
In addition, each level has its own actor and critic functions with the level-based setting. Based on this idea, it is natural to have DDPG-based training steps for each level, {\color{black} while we take the level state of the previous round as input.}

\subsection{Optimize Training with Policy Propagation}\label{sec:propagation}
The training data of the RL model comes from system runtime metrics. For larger levels in the {\FLSM}, training data is sparse because the compaction in a deeper level is exponentially less frequent than at a shallower level. Nonetheless, the size of a deeper level is exponentially larger than a shallower level and hence needs more training data. This gives rise to a crucial question: how to effectively learn the compaction policy at a large level while keeping a fast learning convergence?

To address this issue, we present a {\it policy propagation} method based on a cost analysis. As the query cost analysis is related to the bits-per-key at Bloom filters in each level, discussions are divided divided into two cases:
\begin{itemize}[leftmargin=*]
    \item {\bf Case 1: Uniform Bits-Per-Key:} Every level assigns the same bits-per-key to the Bloom filter. This is the default setting in many key-value stores such as RocksDB~\cite{rocksdb}.
    \item {\bf {Case 2: \Monkey{}} Allocation:} Level $i$ assigns exponentially lower bits-per-key than Level $i+1$. It is shown in~\cite{dayan2017monkey} that this scheme achieves better amortized read performance, and has been adopted in other advanced systems such as Dostoevisky~\cite{dostoevsky2018} and Cosine~\cite{chatterjee2021cosine}.
\end{itemize}

\subsubsection{Propagation for Case 1.}
In this case, the read/write cost ratio is very similar across different levels, for the reason that each level has the same read amplification and write amplification. Therefore, we can use the RL model to learn the policy of Level $1$ and propagate the policy to the other levels. 
Specifically, we employ the per-level model to obtain the optimal policy at Level 1. The compaction policies of other levels remain unchanged throughout this process so as not to interfere with the end-to-end latency which contributes to the reward. Then, we transform the compaction policies at all other levels into the policy we learned at Level 1 with the flexible transition. It should be noted that our previous comments about the potential for independent policy settings at each level still stand. Our earlier mention of independence referred specifically to the design space of the LSM-tree and did not necessarily contradict the possibility of level correlation within an optimal scheme.

\subsubsection{Propagation for Case 2.}
In this case, the read/write cost ratio can be drastically different, and hence the policy at each level can vary. Under the \Monkey\ scheme, if we intend to optimize the expected cost of an operation (lookup or update), there is an optimal policy.  
We mainly focus on analyzing the zero-result case following Dostoevsky~\cite{dostoevsky2018}.
We apply the following lemma to guide us in making appropriate inferences on compaction policies. 

\begin{lemma}
Consider that the three consecutive levels ($i-1$, $i$, $i+1$) in a {\FLSM{}} employ the Monkey scheme. Let $K^*_{i}$ be the optimal merge policy that minimizes the expected cost per operation at Level $i$. We have the following equation:
\vspace{-1mm}
\begin{equation}
\frac{1}{K^*_{i+1}}=\sqrt{{\frac{1}{K^{*2}_i}+T\cdot\left(\frac{1}{K^{*2}_{i}}-\frac{1}{K^{*2}_{i-1}}\right)}}
\label{eq: policy propogation}
\end{equation}
\label{lemma: monkey policy}
\end{lemma}

\vspace{-6mm}
\begin{proof}
We follow the notations in Section~\ref{sec:flsm_prop} (e.g., $K_i$, $B$, $E$, $\gamma$) unless otherwise specified.

\noindent(1) \underline{Lookup cost.}
Let $f_i$ denote the FPR of the Bloom filters in Level $i$. In expectation, the lookup will incur $f_i\cdot K_i$ false positives in Level $i$ in the zero-result case. As each false positive causes one disk page read, the time cost that a lookup causes in Level $i$ is $f_i\cdot I_r\cdot K_i$, where $I_r$ is the average time of a read I/O. Let $c_r$ be the cost of probing the metadata of a sorted run in the main memory. Then, the total lookup cost in Level $i$ is $f_i\cdot I_r \cdot K_i+ c_r\cdot K_i$.

\noindent(2) \underline{Update cost.}
Let $I_w$ denote the average time cost of a write I/O.
According to \cite{idreos2019designcontinuum}, the write amplification of Level $i$ is $T/K_i$, signaling that on average the key-value entry of the update will take part in $T/K_i$ compactions in Level $i$. This entry is read into memory and written into the disk once at each compaction. Therefore, the update incurs an overhead of $\frac{T\cdot E}{B\cdot K_i}\cdot(I_r+I_w)+\frac{T}{K_i}\cdot c_w $, where $c_w$ is the CPU cost a key incurs in compaction (such as merge sorting and space allocation).

Then, the expected time overhead per operation in Level $i$ is
\begin{equation}
{\scriptsize
 \underbrace{f_i \cdot I_r \cdot K_i \cdot \gamma}_{\text{Query I/O cost}}+\underbrace{c_r \cdot K_i \cdot \gamma}_{\text{Query CPU cost}}+\underbrace{\frac{T \cdot E}{B \cdot K_i}\cdot(I_r+I_w)\cdot(1-\gamma)}_{\text{Update I/O cost}}+\underbrace{\frac{T}{K_i} \cdot c_w \cdot (1-\gamma)}_{\text{Update CPU cost}}
 }
 \label{eq: level cost}
\end{equation}

{\Monkey} requires $f_i = T^{i-1}f_1$, where $f_1$ is the FPR of the Bloom filters at the first level. Applying the method of Lagrange multipliers, the time overhead is minimized when
\[
K^{*2}_i = \frac{T \cdot E \cdot(I_r+I_w) \cdot (1-\gamma)+B \cdot T \cdot c_w \cdot (1-\gamma) }{B \cdot f_1 \cdot I_r \cdot T^{i-1} \cdot \gamma+ B \cdot c_r \cdot \gamma}
\]
Or equivalently, 
\[
K^{*2}_i = \frac{X}{Y\cdot T^{i-1} + Z}
\]
where $X= T\cdot E\cdot (I_r+I_w)\cdot (1-\gamma)+T\cdot B \cdot c_w \cdot (1-\gamma)$, $Y=B\cdot f_1\cdot I_r \cdot \gamma$, and $Z=B \cdot c_r \cdot \gamma$. This further gives 

{
\small
\[
\frac{1}{K^{*2}_{i+1}}-\frac{1}{K^{*2}_i}=\frac{Y\cdot (T^i-T^{i-1})}{X}=T\cdot\frac{Y\cdot (T^{i-1}-T^{i-2})}{X}=T\cdot\left(\frac{1}{K^{*2}_{i}}-\frac{1}{K^{*2}_{i-1}}\right),
\]
}
Note that under the Monkey scheme, the optimal policies always satisfy that $K^*_i\leq K^*_{i-1}$, and hence
\[
\frac{1}{K^*_{i+1}}=\sqrt{{\frac{1}{K^{*2}_i}+T\cdot\left(\frac{1}{K^{*2}_{i}}-\frac{1}{K^{*2}_{i-1}}\right)}}
\]
\end{proof}
\vspace{-3mm}

Lemma~\ref{lemma: monkey policy} shows that the optimal policy of two consecutive levels can help us infer the optimal compaction policy of larger levels even if we are unaware of the specific values of the system related parameters $X$, $Y$, and $Z$. Therefore, we let the {\tuner} only tune the compaction policy of Level 1 and Level 2, and infer the compaction policy for all subsequent levels by Lemma~\ref{lemma: monkey policy}. When the inferred policy may not be integers, we round it to the closest valid policy.

\vspace{1mm}
\noindent{\bf Example.}
We assume that the tuning result of Level 1 and Level 2 are 9 and 7, respectively. Then following Equation~\ref{eq: policy propogation}, we set the compaction policy of Level 3 as 
$
{K^*_{3}}=\sqrt{\frac{1}{{\frac{1}{7^2}+10\cdot\left(\frac{1}{7^2}-\frac{1}{9^2}\right)}}}\approx 3
$.
Similarly, Level 4 has
$
{K^*_{4}}=\sqrt{\frac{1}{{\frac{1}{3^2}+10\cdot\left(\frac{1}{3^2}-\frac{1}{7^2}\right)}}}\approx 1
$.

\vspace{1mm}
\noindent\textbf{Additional discussion with non-zero-result lookups.}
The policy propagation still largely remains viable in the presence of non-zero-result lookups. Firstly, according to previous works~\cite{huynh2021endure, chatterjee2021cosine}, the hit rate of non-zero-result lookups for each level is in general proportional to the level capacity, which increases exponentially by a factor of $T$ from the smallest to the largest levels. Consequently, the hit rates for all levels, except the largest level, are relatively insignificant, resulting in a negligible impact on Bloom filters and the trade-off between reads and writes in these levels. Hence, our policy propagation remains applicable in these levels. Secondly, under the Monkey scheme, the FPR of the Bloom filter increases exponentially by a factor of $T$ for each level. As a result, the largest level has an immense FPR, which necessitates an aggressive optimal policy ($K=1$) to remedy excessive read overhead.
For instance, in our case study presented in Figure~\ref{fig:case_study}, the optimal policy for the largest layer is $K=1$.
Even if non-zero-result lookups may partially mitigate the influence of the Bloom filter in the largest level, the need for an aggressive optimal policy very likely remains unchanged.

\section{Related work}

\noindent\textbf{Machine Learning for Database Storage Systems.}
There are extensive studies on tuning storage systems via machine learning. 
Pavlo {\it et al.} \cite{pavlo2017self} and Aken {\it et al.} \cite{van2021inquiry} propose the concept of a self-driving database that makes decisions automatically with machine learning models.
Ottertune~\cite{van2017ottertune} includes the techniques that analyze and flit database knobs with machine learning techniques and further recommend parameter settings by learning from collected metrics.
Qtune~\cite{Qtune2019}, CDBTune~\cite{zhang2019cdbtune}, and CDBTune+~\cite{zhang2021CDBTune+} are RL-based database tuning methods, which adopt DDPG to improve their learning effectiveness in large configuration space.
Bourbon~\cite{dai2020wisckey} also incorporates machine learning into the LSM-tree, however, in a different way from our paper. It proposes to augment the fence pointers in the LSM-tree with the learned index structures~\cite{kraska2018learnedIndex}.
We note that most of the studies are concentrated on tuning SQL databases. To the best of our knowledge, we have the first attempt on applying machine learning to NoSQL key-value database tuning. 

\vspace{1mm}
\noindent\textbf{LSM-tree Optimization}
There are numerous studies on LSM-tree optimizations~\cite{ dostoevsky2018, idreos2019designcontinuum, chatterjee2021cosine, huynh2021endure, dayan2019log, raju2017pebblesdb, dayanspooky, sarkar2020lethe,alkowaileet2019lsm, sarkar2022constructing, zhang2020fpga, thonangi2017log, dayan2021chucky, zhang2018elasticbf, zhu2021reducing, dayan2017monkey,knorr2022proteus,zhang2018surf,ahmad2015compaction, huang2019x, vinccon2018noftl, wang2014efficient,chan2018hashkv, lu2017wisckey,balmau2017triad,golan2015scaling, shetty2013building,luo2020rosetta, wu2015lsmtrie, balmau2019silk, luo2019performance, sears2012blsm, bortnikov2018accordion, kim2020robust, luo2020breaking, absalyamov2018lightweight, ren2017slimdb, yang2020leaper, yu2022treeline}, as we summarized in Section~\ref{sec:intro}. 
Among these studies, some studies~\cite{dayan2017monkey, dostoevsky2018, idreos2019designcontinuum, chatterjee2021cosine, huynh2021endure, dayan2019log} are more related to {\ourmethod}, since they focus on tuning compaction policies as well as {\ourmethod}.
\Monkey~\cite{dayan2017monkey} includes a theoretical cost model for an LSM-tree by estimating the expected I/O cost. Based on the model, \Monkey{} co-tunes the merge policy, the buffer size, and the Bloom filters’ bits-per-key to locate an optimal LSM-tree design with the minimal I/O cost for a given workload.
Dayan and Idreos~\cite{dostoevsky2018} show that existing merge policies (tiering and leveling) are not able to fully trade between read costs and write costs, therefore they propose Fluid LSM-tree that enables hybrid compaction policies. In addition, Dostoevsky further generalizes its theoretical model on hybrid compaction policies.
Chatterjee {\it et al.}~\cite{chatterjee2021cosine} propose a synthetic model named Cosine for key-value stores on the cloud. Moreover, they present an I/O cost model that is aware of workload distribution.
Huynh {\it et al.}~\cite{huynh2021endure} provide a model-based analysis on workloads with uncertainty.
Dayan and Idreos~\cite{dayan2019log} devise a more scalable merge policy for LSM-tree by adjusting the size ratio between adjacency levels.
Compared with these works, our study is the first one that employs machine learning techniques and is a pioneer work that considers real-time transition costs between different tuning settings.

\section{Evaluation}\label{sec:exp}
This section presents experimental studies of {\ourmethod} on various workloads. The result shows that {\ourmethod} exhibits strong robustness in performance across different workloads.

\noindent\textbf{Setup.} We process our experiments on a server with an Intel(R) Xeon(R) Gold 6326 CPU @ 2.90GHz processor, 256GB DDR4 main memory, and 1TB NVMe SSD, running 64-bit Ubuntu 20.04.4 LTS on an ext4
partition.

\vspace{1mm}
\noindent\textbf{Implementation.} 
We implement {\ourmethod}'s {\FLSM} on top of Facebook's RocksDB \cite{rocksdb}, a widely used LSM tree-based key-value storage system. 
Similar to previous works \cite{dostoevsky2018,chatterjee2021cosine, dayan2019log}, we extend the compactor API which RocksDB provides to allow hybrid compaction policies across levels.
We also implement {\Monkey} \cite{dayan2017monkey} by enabling RocksDB to assign different bits-per-keys of the Bloom filters across all levels.
Following previous works \cite{dostoevsky2018,huynh2021endure}, we enable direct I/Os for read and write. We set the memory buffer size to 2MB, the size ratio to 10, and the default Bloom filter bits-per-key to 8. For RocksDB with {\Monkey} scheme, we lower bits-per-key to 4 since in this case Monkey exploits Bloom filters more effectively. We use the internal counters of RocksDB to acquire necessary statistics for states and measure the level-based latency.
We implement the {\RLmodel} model with PyTorch. In our experiments, we employ a three-layer fully-connected neural network with 128 neurons per layer using ReLU activation functions for both critic and actor. 
We let the end-to-end latency and by-level latency to be equally contributed to the reward by setting parameter $\alpha$ to $1/2$.

Our previous discussions are based on the assumption that the compaction is conducted per level at a time, in line with many previous works~\cite{dayan2017monkey,dostoevsky2018,dayan2019log,huynh2021endure}.
In RocksDB, however, compaction is performed at the granularity of files. A sorted run in these systems is composed of a few disk files. During compaction, a small set of files with overlapping key ranges in each level are merged. File compaction may reduce the actual size of a sealed run, which will be reactivated, and ultimately leads to multiple active runs in one level. Although the compaction granularity is different, a similar transition method can still be applied. To increase the compaction policy $K_i$ of Level $i$ by one, we reduce the capacity of all existing runs in the level from $\frac{C_i}{K_i}$ to $\frac{C_i}{K_i+1}$, where $C_i$ is the level capacity of Level $i$, and initialize an empty run as the active run with the same capacity. To decrease $K_i$ by one, we seal one of the runs in the level and increase the capacity of other sequences from $\frac{C_i}{K_i}$ to $\frac{C_i}{K_i-1}$. To avoid an additional cost and a performance slump, the sealed run does not trigger any compaction.
When other runs in the level trigger a file compaction, overlapping files in the sealed run will also be involved in this compaction. Therefore, the sealed run would gradually get merged through compactions triggered by other runs in the level. 
In this way, a {\FLSM} transition still takes effect immediately and incurs zero immediate transition cost. 

\vspace{1mm}
\noindent\textbf{Baselines.} We compare {\ourmethod} with three typical compaction policy settings in Fluid LSM-tree~\cite{dostoevsky2018}, including aggressive compaction ($K=1$), lazy compaction ($K=10$), and moderate setting ($K=5$). We name them as {\aggr}, {\lazy} and {\mode}, respectively. We also compare {\ourmethod} with the state-of-the-art hybrid compaction policy {\lazyl}. For fairness, we only compare {\lazyl} under {\Monkey} scheme, following the setting of Dostoevisky~\cite{dostoevsky2018}.
{\color{black}
In addition, we also include heuristic-based baselines on top of our {\FLSM}, which greedily selects a more read or write friendly compaction policy for each level if the workload in the level moves towards the respective direction.
}

\vspace{1mm}
\noindent\textbf{Experiment Design.}
By default, we initially bulk load the empty database with 100 million random sampled key-value entries. Each key-value entry consists of a key of 128 bytes and a value of 896 bytes.
The mission size is set to 50000 operations.
{\color{black}
The setting is determined  to ensure a reasonably consistent workload throughout a mission, while also involving sufficient operations in a mission to gather the necessary information for reinforcement learning. 
}
We measure the performance of the database with workloads of 100 million operations, which are divided into 2000 missions. Each operation can be {\it lookup} or {\it update}, which consists of uniformly and randomly distributed keys and values.
Our evaluations are performed with various ratios between lookups and updates. For {\ourmethod}, we initiate the {\FLSM} with leveling, which is the default compaction policy in RocksDB.

We evaluate {\ourmethod} under both the uniform Bloom filter scheme and the {\Monkey} Bloom filter scheme. In virtue of policy propagation, we only need to tune the compaction policy Level 1 under the uniform scheme and propagate it to other larger levels. Under the Monkey scheme, we tune Level 1 and Level 2 successively with the {\RLmodel}. \textcolor{black}{When the compaction policies of Level 1 and Level 2 stay stable and the RL model converges, we finish the tuning and transfer the policy to larger levels with Equation~\ref{eq: policy propogation}}. 

\begin{figure*}[t]
  \centering 
  \makebox[0pt][c]{
    \includegraphics[width=14cm]{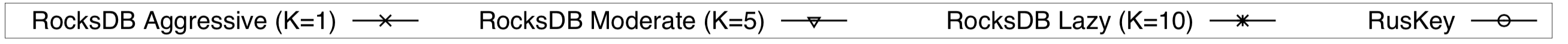}
    }
    \vspace{-3mm}
\begin{minipage}[c][0mm][t]{\linewidth}
    \vspace{0mm}
\makebox[\linewidth][c]{
}
\end{minipage}
  \makebox[0pt][c]{
    \hspace{1.2mm}
    \begin{overpic}[width=4.65cm]{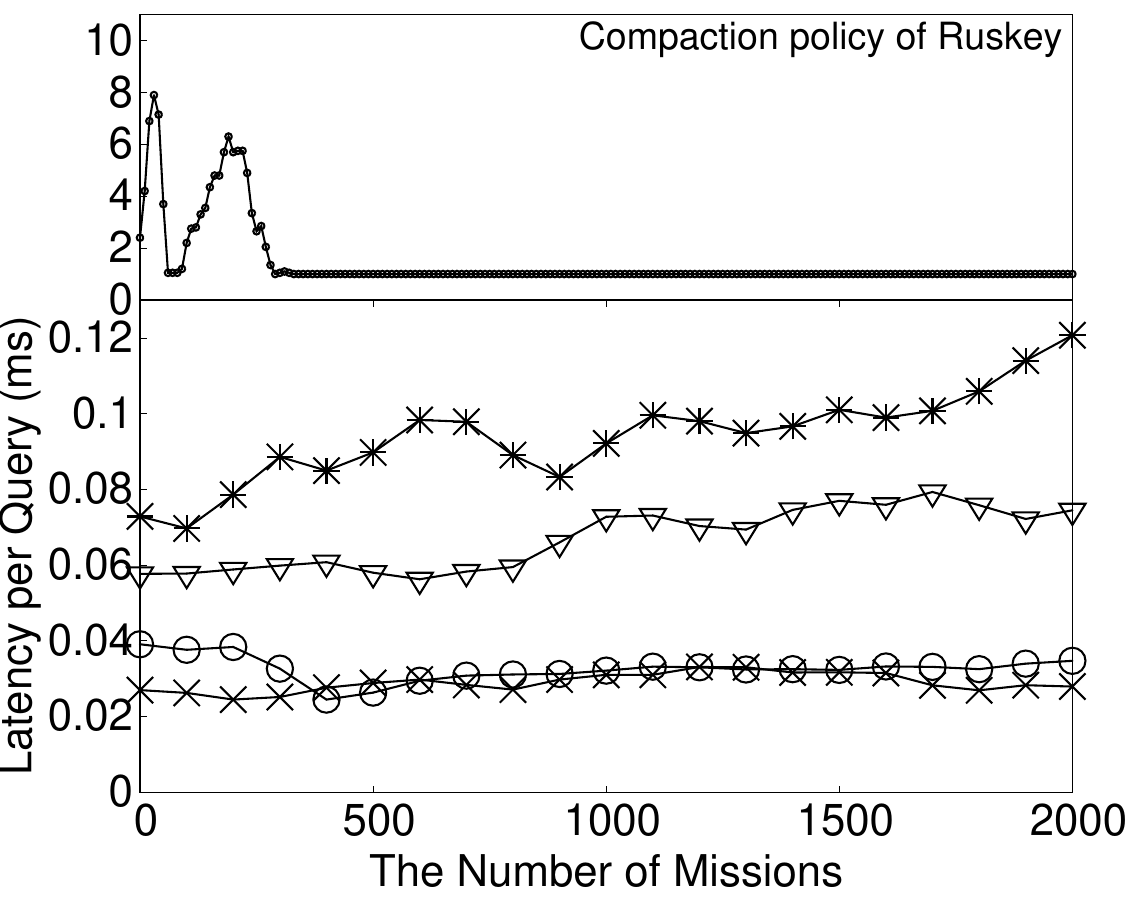}\put(14, 48){\small\color{black}{(a)~Read-heavy}}\end{overpic}
    \hspace{0.4mm}
    \begin{overpic}[width=4.65cm]{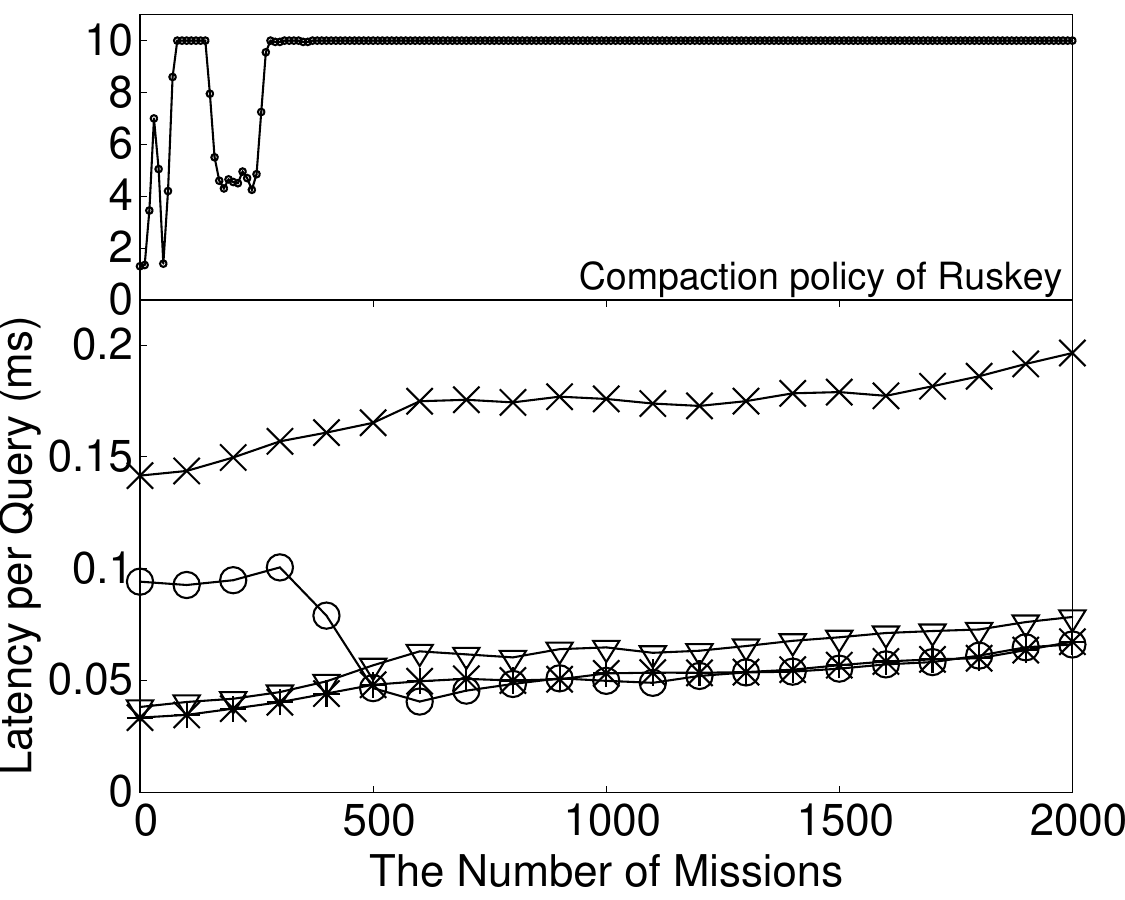}\put(14, 48){\small\color{black}{(b)~Write-heavy}}\end{overpic}
    \hspace{0.4mm}
    \begin{overpic}[width=4.65cm]{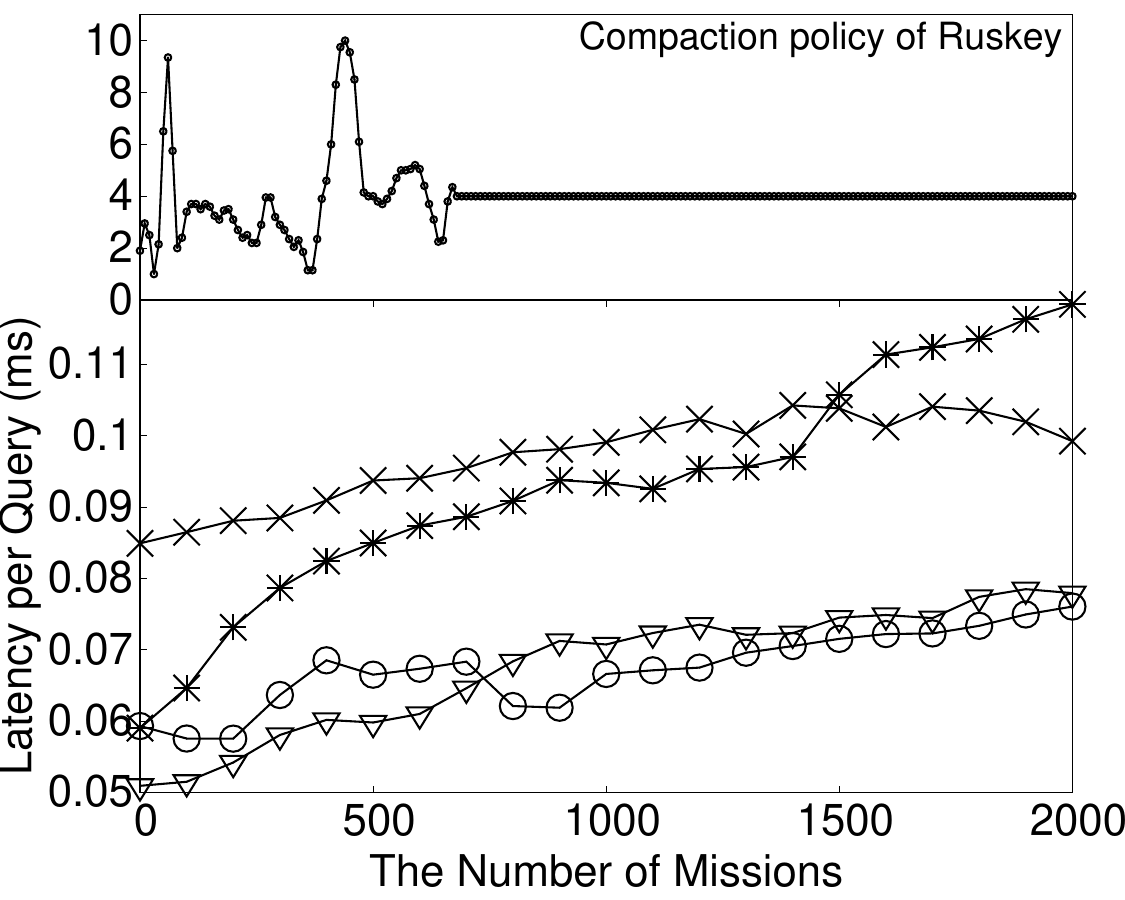}\put(14, 48){\small\color{black}{(c)~Balanced}}\end{overpic}
    \hspace{0.4mm}
}
\vspace{-2mm}
  \caption{{\ourmethod} self-navigates to the optimal design on static workloads. 
  \vspace{2mm}
  }
  \label{fig: latency no monkey}
\end{figure*}

\vspace{1mm}
\noindent\textbf{{\ourmethod} self-navigates to optimal design on static workloads.} 
In Figure~\ref{fig: latency no monkey}, we compare the performance of {\ourmethod} with the baselines under various workloads, which include read-heavy workload with 90\% lookups, write-heavy workload with 10\% lookups, and read-write-balanced workload with 50\% lookups. 
Each of these workloads includes 100 million operations which are divided into 2000 missions.

The result shows that {\ourmethod} achieves near-optimal performance across all workloads.
In contrast, the baselines perform sub-optimally on at least one workload. For example, in the read-heavy workload (Figure~\ref{fig: latency no monkey} (a)), {\aggr} is the most effective because it maintains a low read cost by merging sorted runs aggressively {that promises low read amplification}. However, its performance degrades with the write-heavy workload (Figure~\ref{fig: latency no monkey} (b)) because the aggressive merging incurs significant additional write amplification. {\mode} achieves moderate performance on each of the workloads. It performs worse than aggressive policy on the read-heavy workload and worse than lazy policy on the write-heavy workload. Meanwhile, its performance is relatively better than the other two baselines on the balanced workload. To explain, the balanced workload 
contains an equal quantity of lookups and updates. {\aggr} and {\lazy} are respectively much slower on update and lookup operations. In contrast, {\mode} processes both lookups and updates with reasonable efficiency. Nevertheless, the end-to-end performance of moderate policy is still lower than {\ourmethod} on the balanced workload when {\ourmethod} finishes the tuning and reaches a sweet trade-off between lookups and updates.

Overall, {\ourmethod} has robust performance across different workloads as it can adjust its compaction policies to accommodate different workload compositions. 
The sub-figures at the top of Figure~\ref{fig: latency no monkey} show the compaction policy of {\ourmethod} tuned by the RL model {\RLmodel} during workload processing. 
Through tracking the tuning strategy of {\RLmodel}, we find that {\RLmodel} selects an aggressive compaction policy with a read-heavy workload and a lazy compaction policy with a write-heavy workload. 
In addition, as shown in Figure~\ref{fig: latency no monkey} (c), {\RLmodel} selects an intermediate policy $K=4$ on the balanced workload, which dominates all baselines after around 600 missions. 
This is because the baselines are sub-optimal on the balanced workload while {\ourmethod} self-drives to the optimal compaction policy through the {\RLmodel} model.
This wise strategy is enabled by integrating the {\FLSM} with the deep RL model {\RLmodel}. Note that there are still minor gaps between {\ourmethod} and the respective optimal baselines on read-heavy and write-heavy workloads because {\ourmethod} requires a small number of missions to conduct the online training in order to find the optimal compaction policy. As shown in the upper sub-figures of Figure~\ref{fig: latency no monkey}, it takes {\RLmodel} around 300 missions to explore and navigate to the optimal setting with the read-heavy and write-heavy workloads. Tuning on the balanced workload takes longer, which is around 600 missions.

\vspace{1mm}
\noindent\textbf{{\ourmethod} performs better on dynamic workloads.}
In this experiment, we evaluate {\ourmethod} and the baselines with a comprehensive dynamic workload. The workload is divided into five sessions with different proportions of lookups and updates. They are in the order of read-heavy (10\% update), balanced (50\% update), write-heavy (90\% update), write-inclined (70\% update), and read-inclined (30\% update). Each section includes 50 million operations which are divided into 1000 missions with 50000 operations for each.

Figure~\ref{fig: latency with dynamic workload} presents the latency of {\ourmethod} and the baselines as well as the compaction policy that {\ourmethod} chooses. All baselines perform sub-optimally on at least one of the sessions. 
In the meantime, the latency of {\ourmethod} is close-to-optimal across all sessions, benefiting from a successful tuning through {\RLmodel} and {\FLSM}'s ability to flexibly transform the compaction policy. Specifically, 
{\ourmethod} adopts compaction policy settings of $K=1$, $K=5$, $K=10$, $K=6$, $K=2$ on the read-heavy, balanced, write-heavy, write-incline and read-inclined sessions, respectively.

Across all workload sessions, {\ourmethod} achieves up to 4 times better performance than {\aggr} (on the write-heavy session), up to 2.4 times better performance than {\mode} (on the read-heavy session), and up to 3.5 times better performance than {\lazy} (on the read-heavy session). Note that in Figure~\ref{fig: latency with dynamic workload}, the latency of {\ourmethod} rises slightly at the beginning of each workload session and then drops rapidly. 
The reason behind this is that {\RLmodel} requires a short retraining under a new workload setting. 

To further show the robustness of {\ourmethod}, we compute the performance ranking of all methods across all five workload sessions {\color{black} after tuning {\ourmethod}}, which is summarized in Table~\ref{table:ranking}. The performance is compared by the average time cost per operation after the RL model is converged in each session. The result demonstrates that {\ourmethod} ranks first for four workloads and ranks second for the remaining one. Overall, it achieves a high average ranking of 1.2, demonstrating the strong robustness of {\ourmethod}. 

\begin{figure*}[t]
\vspace{0mm}
    \centering  
    \hspace{5.5mm}
    \includegraphics[width=14cm]{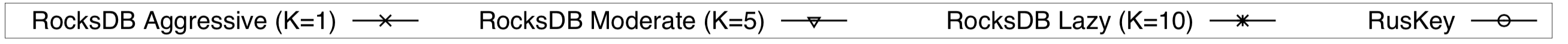}
    \vspace{-8mm}
\end{figure*}
\begin{figure*}[t]
  \centering 
  \makebox[0pt][c]{
    \includegraphics[width=1.03\textwidth]{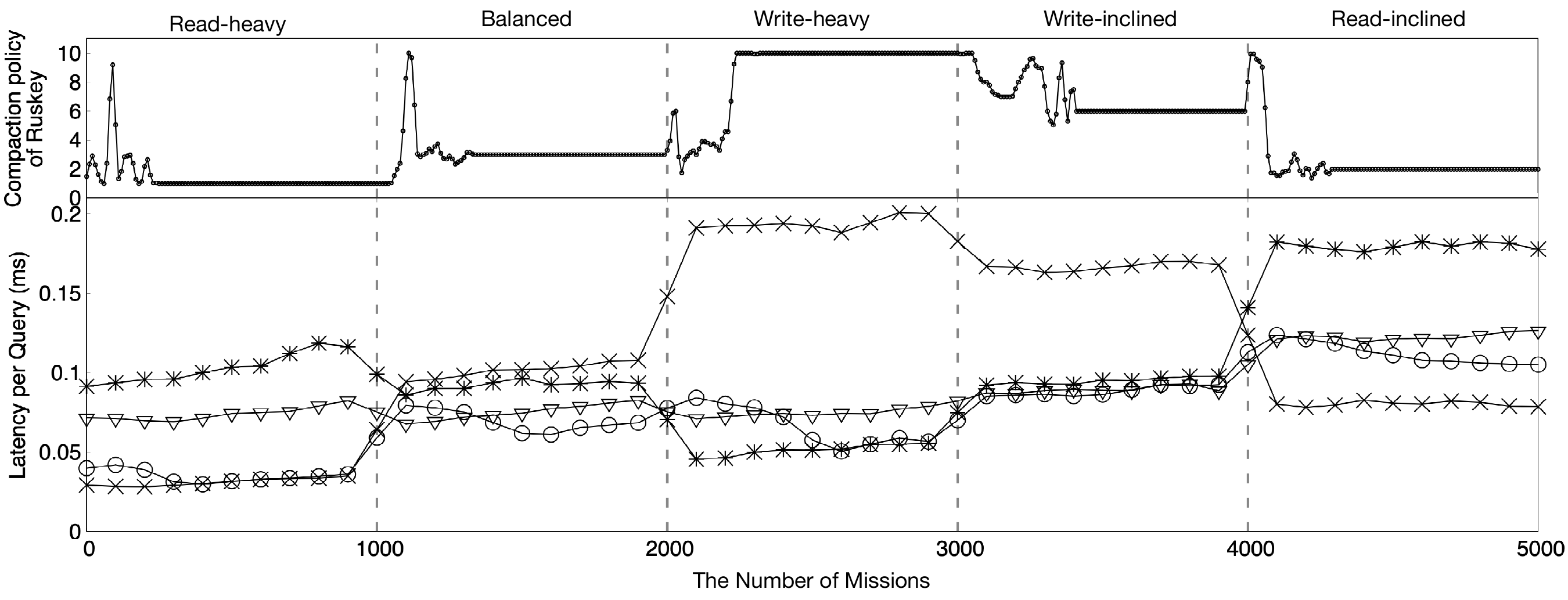}
}
\vspace{-3mm}
  \caption{{\ourmethod} outperforms other baselines on dynamic workloads through self-tuning and transforming its compaction policy according to the change of workload.}
\vspace{0mm}
  \label{fig: latency with dynamic workload}
\end{figure*}

{
\small
\begin{table}[t]
\renewcommand\arraystretch{1.1}
   \addtolength{\tabcolsep}{1.2pt}
\scalebox{0.85}{
\begin{tabular}{|l|ccccc|c|}
\hline
\multirow{2}{*}{{\bf Method}}       & \multicolumn{5}{c|}{{\bf Performance Ranking}}                                                                                                                          & \multicolumn{1}{l|}{\multirow{2}{*}{\bf Avg. Ranking}} \\ \cline{2-6}
                              & \multicolumn{1}{l|}{Read-H} & \multicolumn{1}{l|}{Balanced}    & \multicolumn{1}{l|}{Write-H} & \multicolumn{1}{l|}{Write-I} & \multicolumn{1}{l|}{Read-I} & \multicolumn{1}{l|}{}                                 \\ \hline
K=1                           & \multicolumn{1}{c|}{\textbf{1}} & \multicolumn{1}{c|}{4}          & \multicolumn{1}{c|}{4}           & \multicolumn{1}{c|}{4}             & \textbf{1}                        & 2.8                                                   \\ \hline
K=5                           & \multicolumn{1}{c|}{3}          & \multicolumn{1}{c|}{2}          & \multicolumn{1}{c|}{3}           & \multicolumn{1}{c|}{\textbf{1}}    & 3                                 & 2.4                                                   \\ \hline
K=10                          & \multicolumn{1}{c|}{4}          & \multicolumn{1}{c|}{3}          & \multicolumn{1}{c|}{\textbf{1}}  & \multicolumn{1}{c|}{3}             & 4                                 & 3                                                     \\ \hline
{\ourmethod} & \multicolumn{1}{c|}{\textbf{1}} & \multicolumn{1}{c|}{\textbf{1}} & \multicolumn{1}{c|}{\textbf{1}}  & \multicolumn{1}{c|}{\textbf{1}}    & 2                                 & 1.2                                                   \\ \hline
\end{tabular}
}
\vspace{2mm}
\caption{{\ourmethod} ranks higher in performance on average. {\color{black} ``H'' stands for ``heavy'' and ``I'' stands for ``inclined''.} }\label{table:ranking}
\vspace{0mm}
\end{table}
}

\vspace{1mm}
\noindent\textbf{{\ourmethod} is effective under the Monkey scheme.}
In Figure~\ref{fig: latency with monkey}, we compare {\ourmethod} with the baselines under the {\Monkey} scheme, with the same workload setting as Figure~\ref{fig: latency no monkey}. In addition to the three basic baselines, we also compare with {\lazyl}~\cite{dostoevsky2018}, which is one of the state-of-the-art compaction policies designed for the {\Monkey} scheme. The result is similar to Figure~\ref{fig: latency no monkey}, showing that {\ourmethod} achieves near-optimal performance across various workloads under the Monkey scheme as well. Moreover, {\lazyl} also reaches near-optimal across all workloads, whereas {\ourmethod} performs better than {\lazyl} on every workload. Specifically, {\ourmethod} performs significantly better than {\lazyl} and other baselines on the balanced workload. The reason is that {\ourmethod} is able to adopt novel policy settings through policy propagation. We will explain this in detail with a subsequent experiment.

\begin{figure*}[t]
  \centering 
  \makebox[0pt][c]{
  \includegraphics[width=14cm]{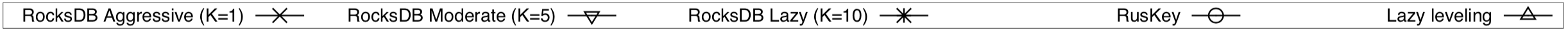}
  }
  \vspace{-4mm}
  \begin{minipage}[c][0mm][t]{\linewidth}
    \vspace{0mm}
\makebox[\linewidth][c]{
}
\end{minipage}
  \makebox[0pt][c]{
    \hspace{1.2mm}
    \begin{overpic}[width=4.5cm]{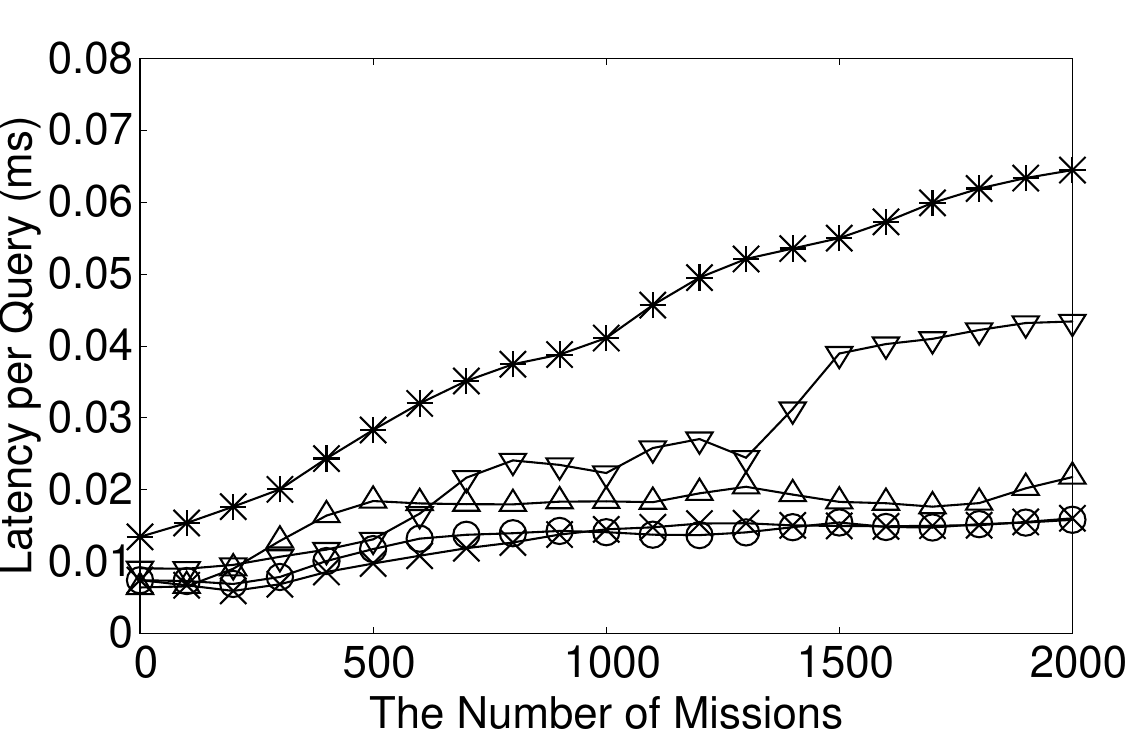}\put(15,55){\small\color{black}{(a)~Read-heavy}}\end{overpic}
    \hspace{0.3mm}
    \begin{overpic}[width=4.5cm]{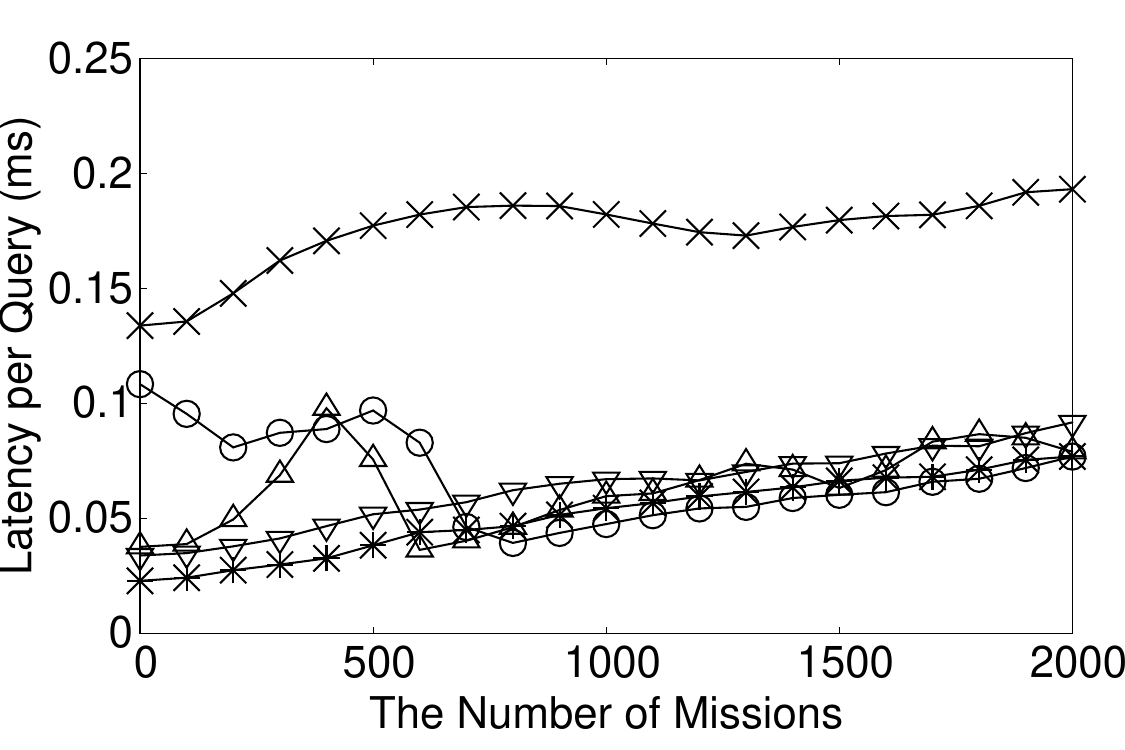}\put(15,55){\small\color{black}{(b)~Write-heavy}}\end{overpic}
    \hspace{0.3mm}
    \begin{overpic}[width=4.5cm]{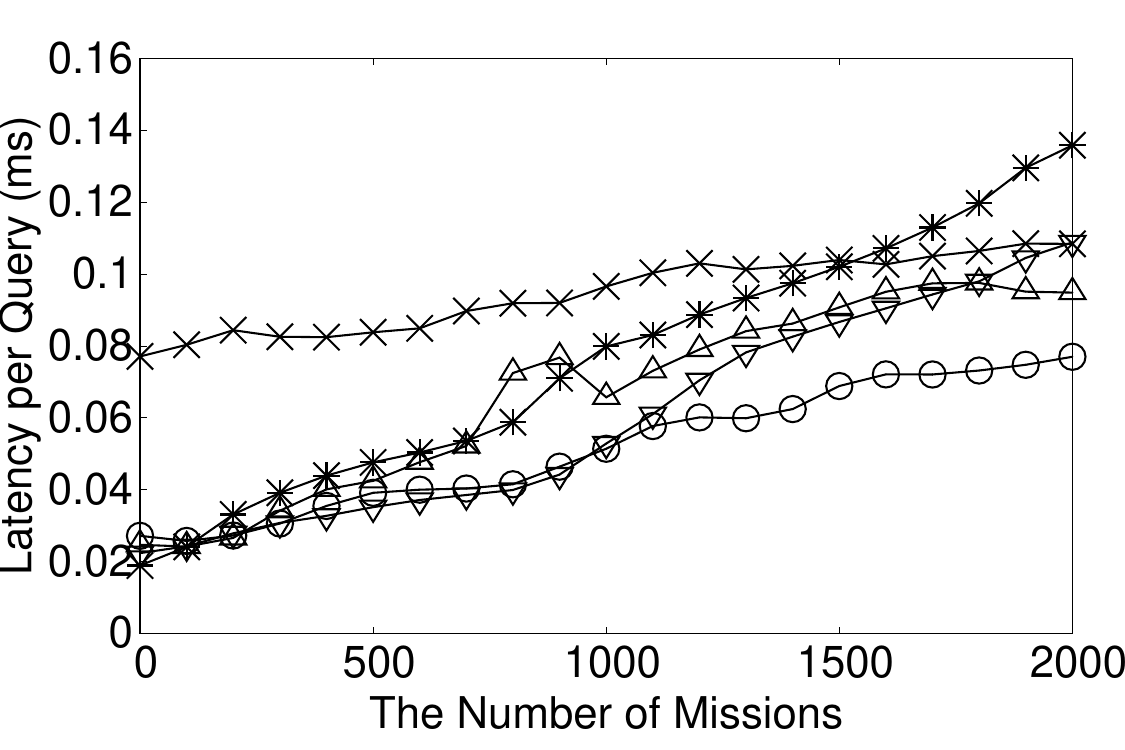}\put(15,55){\small\color{black}{(c)~Balanced}}\end{overpic}
    \hspace{0.3mm}
}
\vspace{-2mm}
  \caption{{\ourmethod} is effective under the {\Monkey} scheme.}
\vspace{0mm}
  \label{fig: latency with monkey}
\end{figure*}

\begin{figure*}[t]
  \centering
  \hspace{0mm}
    \includegraphics[width=0.98\textwidth]{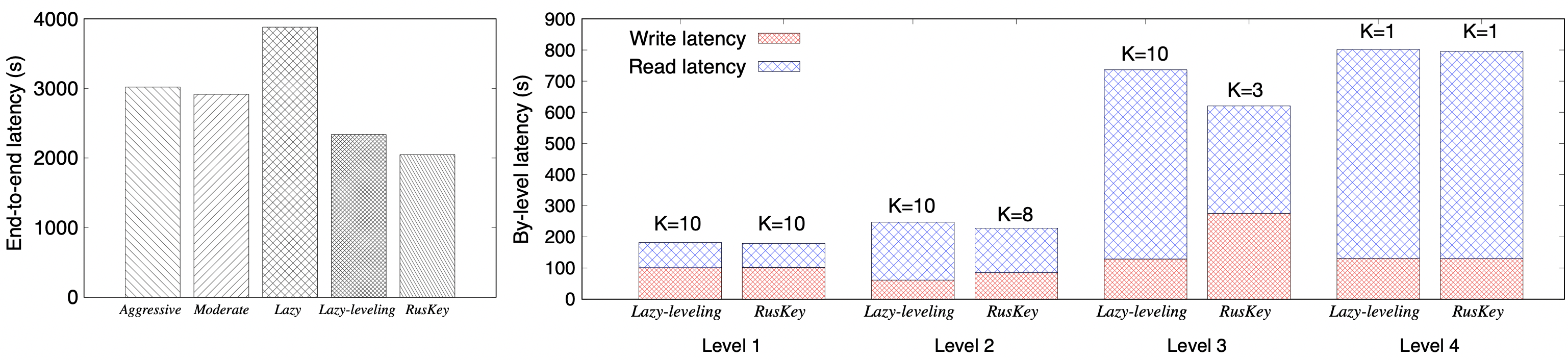}
\vspace{-2mm}
  \caption{{\ourmethod} adopts novel policy settings compared with previous designs.}\label{fig:case_study}
\vspace{2mm}
\end{figure*}

\vspace{1mm}
\noindent\textbf{{\ourmethod} adopts novel policy settings compared with previous designs.}
{
\color{black}
We compare the compaction policy setting of {\ourmethod} and the state-of-the-art design {\lazyl} under the read-write-balanced workload.
The experiment is conducted on a workload of 25 million operations under the {\Monkey} scheme after {\ourmethod} finishes self-tuning.
We report the end-to-end latency (i.e., the running time of processing all the operations) in Figure~\ref{fig:case_study} Left and the per-level latency in Figure~\ref{fig:case_study} Right. 
}
Under this case, {\ourmethod} adopts different compaction policies at each level. It sets an aggressive policy at the first level. As the level goes deeper, the policy becomes lazier. 
This matches the intuition of {\lazyl}, that is, compactions in smaller levels are less needed than in smaller levels with the {\Monkey} scheme, because the read cost is lower in smaller levels. 
The superiority of {\ourmethod} is that {\ourmethod} self-tunes its compaction policy in a more elaborate way than {\lazyl} and gains more benefits from this intuition.

{
\color{black}
\vspace{1mm}
\noindent\textbf{{\ourmethod} remains effective when for YCSB workloads.}
We evaluate {\ourmethod} under the  YCSB standard benchmarks and report the results in Figure~\ref{fig: latency with YCSB}. We use the default Zipfian distribution, in which the update frequency and access frequency of keys follow the power law. As shown in Figure~\ref{fig: latency with YCSB} (a), (b), and (c), the results on YCSB are quite similar in trends to the results on uniform workloads, if we set the composition of workloads the same as the previous experiments. These results further verify that {\ourmethod} delivers near-optimal performance across diverse workloads on the YCSB benchmark. Furthermore, we evaluate YCSB benchmark with a workload that contains 50\% range lookups and 50\% updates, and the results are shown in Figure~\ref{fig: latency with YCSB} (d). Among all baselines, {\aggr} achieves the lowest latency, and the performance of {\ourmethod} is on par with that of {\aggr}. This demonstrates that {\ourmethod} remains high effectiveness even with range lookups. 
}

\begin{figure*}[t]
  \centering 
  \makebox[0pt][c]{
  \centering  
    \hspace{1.3mm}
  \includegraphics[width=9.08cm]{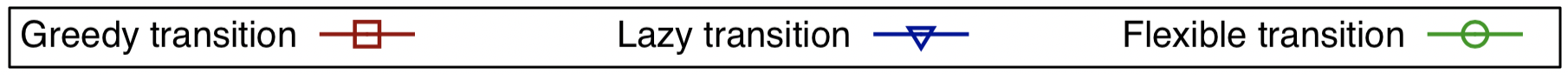}
  }

  \makebox[0pt][c]{
    \hspace{0mm}
    \begin{overpic}[width=4.75cm]{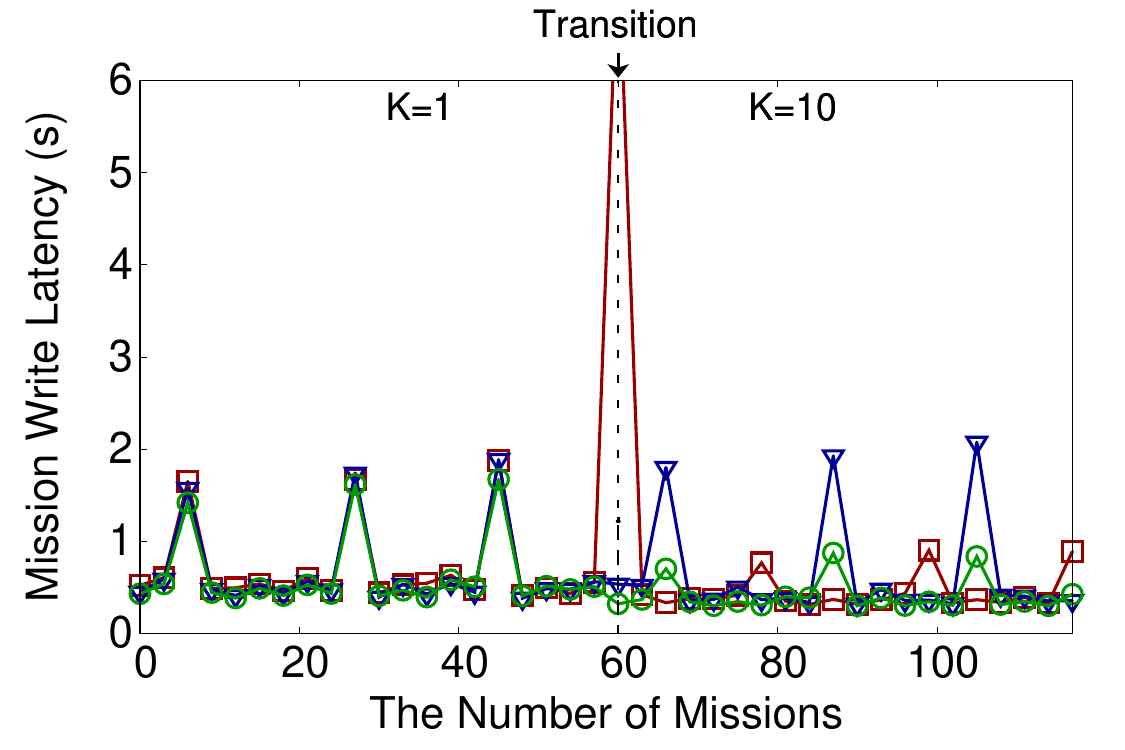}\put(30,-8){\small\color{black}{(a)~Write latency}}\end{overpic}
    \hspace{1mm}
    \begin{overpic}[width=4.75cm]{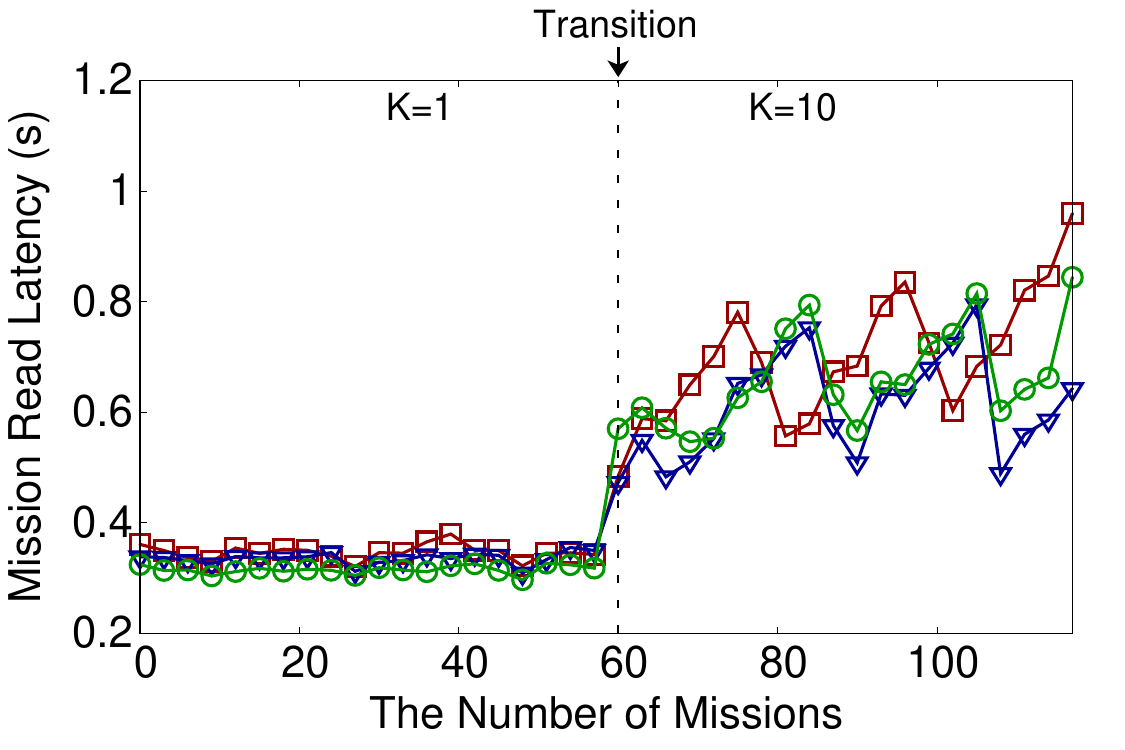}\put(30,-8){\small\color{black}{(b)~Read latency}}\end{overpic}
    \hspace{0.4mm}
}
\vspace{3mm}
  \caption{Flexible transition incurs no immediate transition cost and delay.}
  \label{fig: transition exp}
  \vspace{2mm}
\end{figure*}

\begin{figure*}[t]
    \centering  
    \vspace{-4mm}
    \hspace{4.5mm}
    \includegraphics[width=14cm]{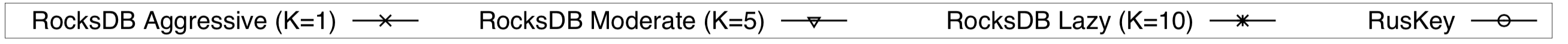}
\end{figure*}

\begin{figure*}[t]
\vspace{-4mm}
  \centering 
  \makebox[0pt][c]{
    \begin{overpic}[width=4.8cm]{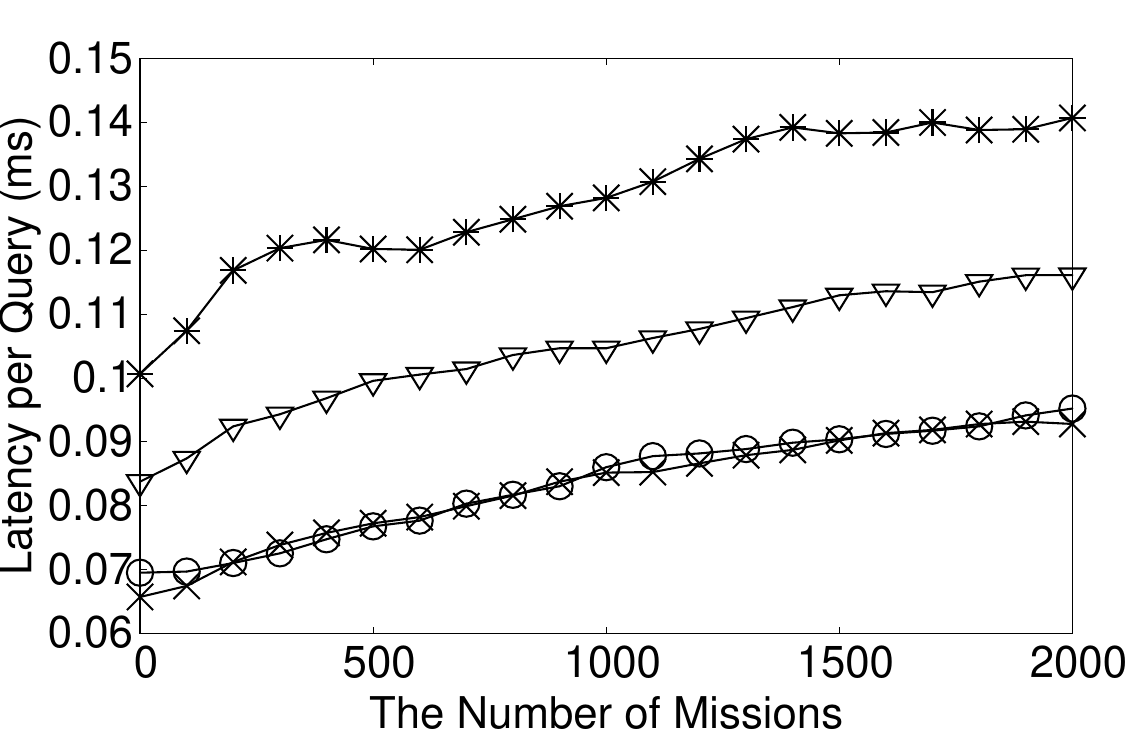}\put(15,55){\color{black}\small{(a)~Read-heavy}}\end{overpic}
    \hspace{1mm}
    \begin{overpic}[width=4.8cm]{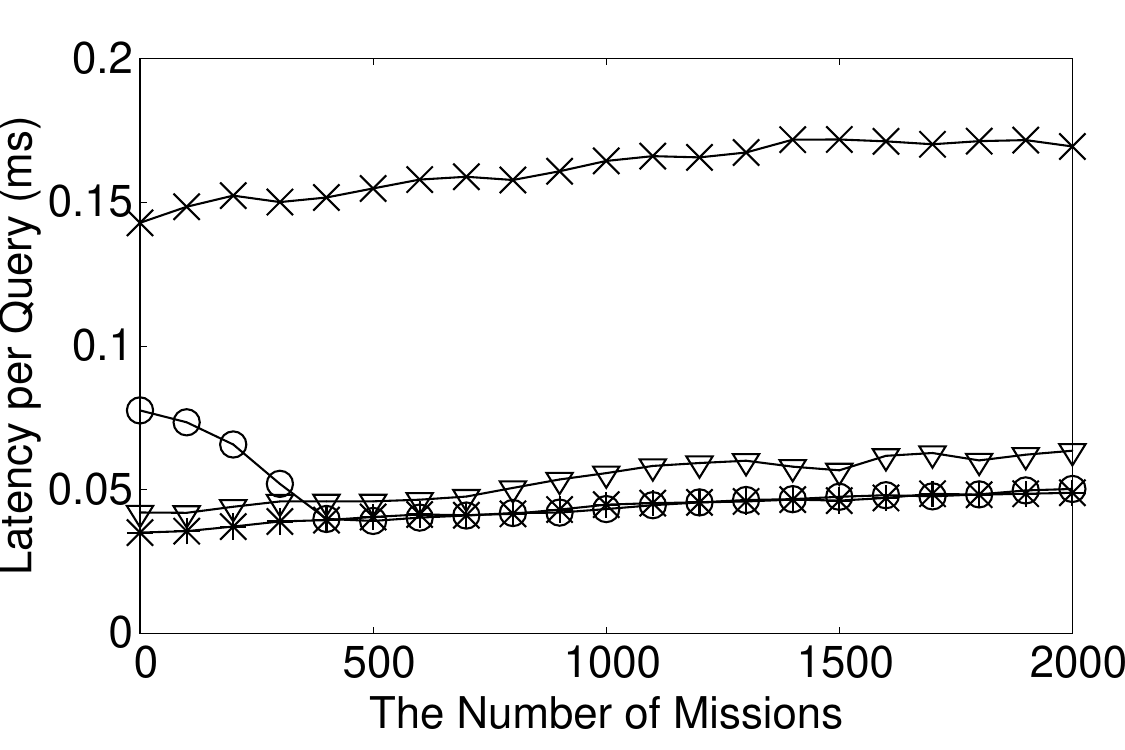}\put(15,55){\color{black}\small{(b)~Write-heavy}}\end{overpic}
    }
    \vspace{-4mm}
    \begin{minipage}[c][0mm][t]{\linewidth}
    \vspace{0mm}
\makebox[\linewidth][c]{
}
\end{minipage}
\makebox[0pt][c]{
\hspace{0.15mm}
    \begin{overpic}[width=4.8cm]{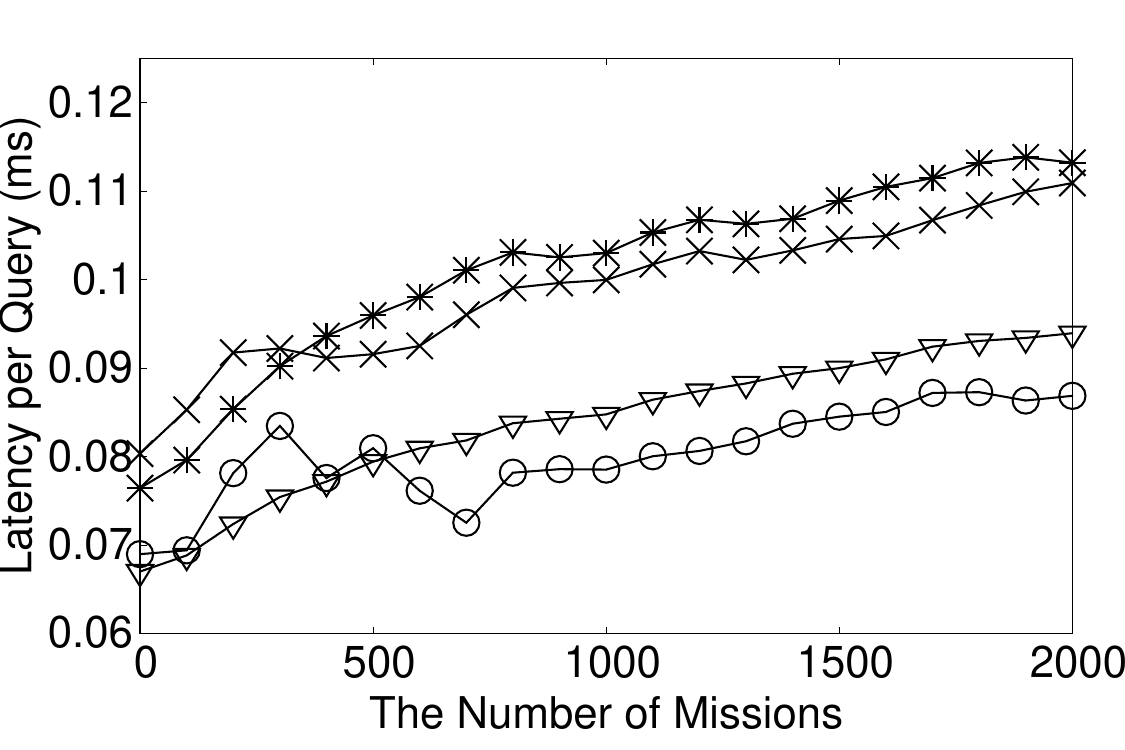}\put(15,55){\color{black}\small{(c)~Balanced}}\end{overpic}
    \hspace{1mm}
    \begin{overpic}[width=4.8cm]{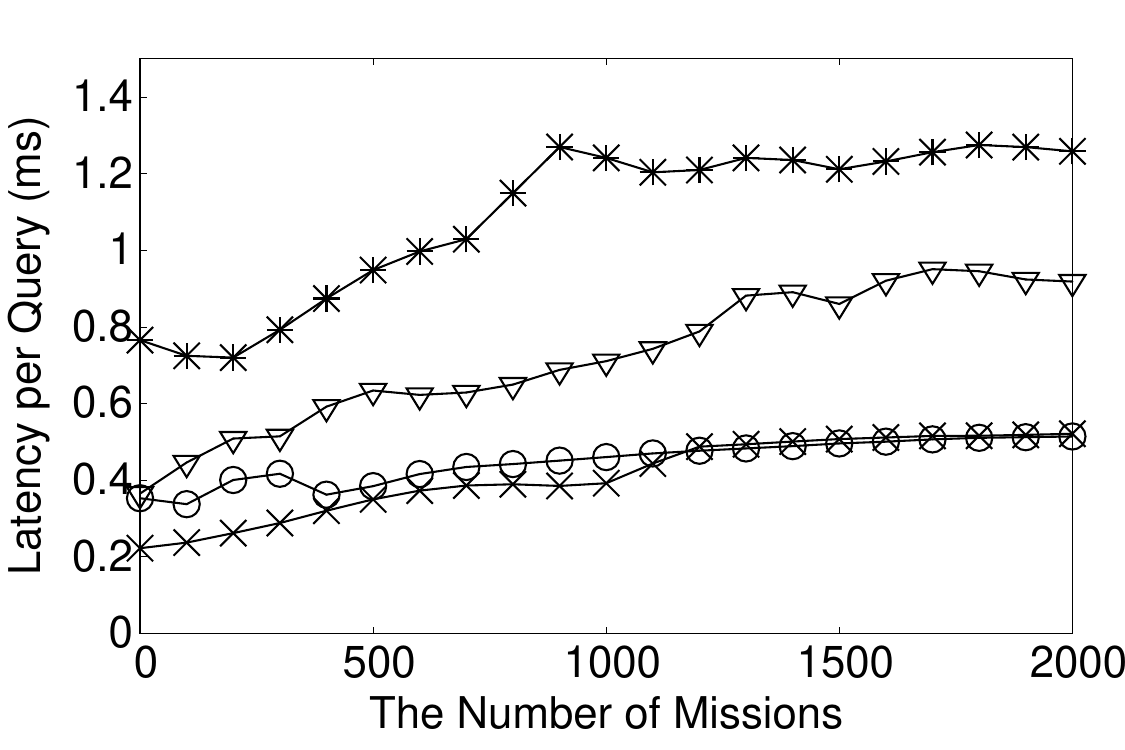}\put(15,55){\color{black}\small{(d)~Range}}\end{overpic}
    }
\vspace{-2mm}
  \caption{\color{black} {\ourmethod} is robust on YCSB benchmarks and can self-tune to a near-optimal policy.}
\vspace{2mm}
  \label{fig: latency with YCSB}
\end{figure*}

\vspace{1mm}
\noindent\textbf{Flexible transition performs better than other transition methods.}
In Figure~\ref{fig: transition exp}, we evaluate the greedy, lazy, and flexible transitions using a micro-benchmark with compaction granularity of levels.
We employ a read-write-balanced workload (50\% lookups and 50\% updates) of 120 missions, while each mission contains 1 million operations. 
Each level's initial compaction policy in the {\FLSM} is 1.
The compaction policy is transformed from 1 to 10 midway through the workload. Specifically, the compaction policy transition happens at the 60th mission, which is represented by the dotted lines in Figure 10.
In Figures~\ref{fig: transition exp} (a) and~\ref{fig: transition exp} (b), the write latency and read latency for each mission using various transition methods are depicted, respectively.
The spikes of write latency in Figures~\ref{fig: transition exp} (a) are caused by compactions with the level granularity.
Figure~\ref{fig: transition exp} (a) demonstrates that when a greedy transition occurs, it results in a very high write latency spike since changing the compaction policies at every level necessitates compacting all data in each level to the next level.
With a greedy transition and a flexible transition, the write latency spikes brought on by compactions are significantly lower after the policy is changed. 
The reason is that the new compaction policy of $K=10$ incurs less write amplification, so the cost of each compaction, i.e. the spikes, would be lower. However, with a lazy transition, the height of spikes stays the same before and after the transition, which means that the compaction policies in larger levels still remain unchanged, because the transition is delayed until the level is completely filled. Moreover, we can find similar results in Figures~\ref{fig: transition exp} (b). The read latencies of the greedy and flexible transition are higher than that of the lazy transition after the transition happens, which also reflects the delay of the lazy transition. {
\color{black}
In terms of the end-to-end latency, the greedy transition, lazy transition and flexible transition use 51s, 44s, and 40s, respectively. 
}
In general, Figure~\ref{fig: transition exp} indicates that the flexible transition incurs almost-free immediate transition cost and has no delay, demonstrating its superiority. 

\begin{figure*}
\vspace{0mm}
  \makebox[0pt][c]{
    \hspace{-12mm}
        \begin{minipage}[h]{0.82\textwidth}
            \centering
            \includegraphics[width=0.99\textwidth]{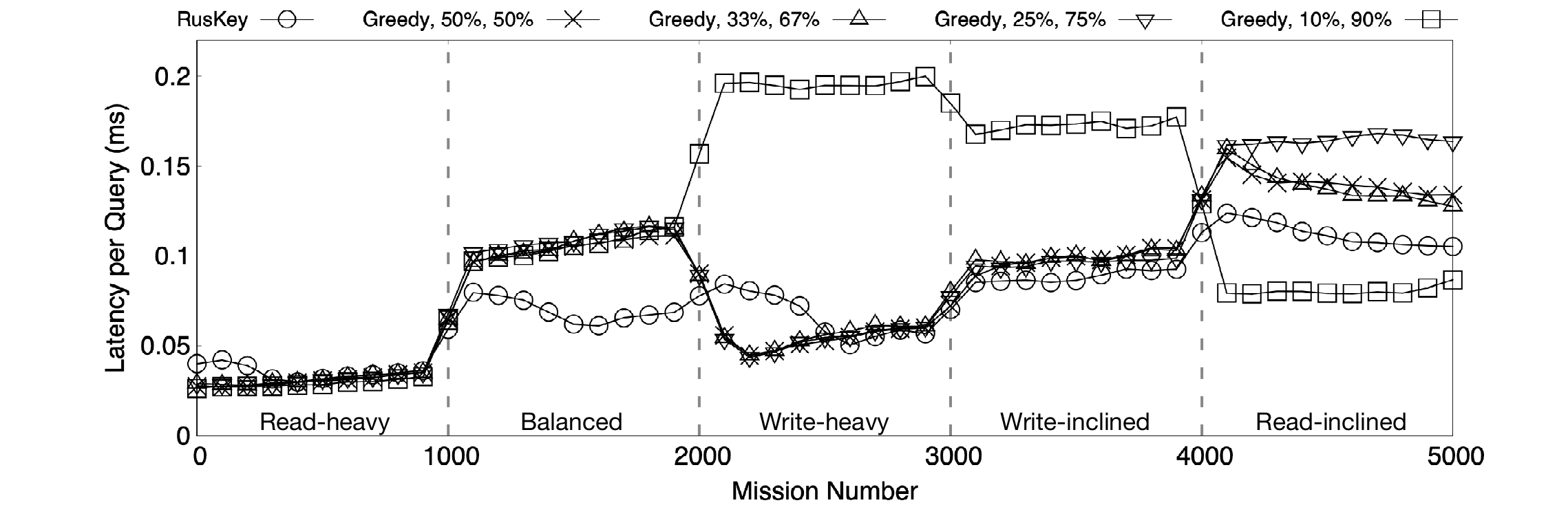}   
        \end{minipage}
        \hspace{-7mm}
        \begin{minipage}[h]{0.25\textwidth}
        \vspace{-3mm}
            \centering
            \renewcommand\arraystretch{1.24}
   \addtolength{\tabcolsep}{-2.5pt}
\color{black}
\scalebox{0.66}{
  \begin{tabu}{|cl|c|}
\hline
\multicolumn{2}{|c|}{\textbf{Method}}                                                                                         & \textbf{Rank} \\ \hline
\multicolumn{2}{|c|}{\ \ \ \ {\ourmethod}}                                                                           & 1.2           \\ \hline
\multicolumn{1}{|c|}{\multirow{4}{*}{\begin{tabular}[c]{@{}c@{}}Symmetric\\ Thresholds\end{tabular}}} & Greedy, 50\%, 50\% & 1.8           \\ \cline{2-3} 
\multicolumn{1}{|c|}{}                                                                                   & Greedy, 33\%, 67\% & 1.8           \\ \cline{2-3} 
\multicolumn{1}{|c|}{}                                                                                   & Greedy, 25\%, 75\% & 2.4           \\ \cline{2-3} 
\multicolumn{1}{|c|}{}                                                                                   & Greedy, 10\%, 90\% & 3.2           \\ \hline
\multicolumn{1}{|c|}{\multirow{2}{*}{\begin{tabular}[c]{@{}c@{}}Biased\\ Thresholds\end{tabular}}}    & Greedy, 25\%, 50\% & 1.8           \\ \cline{2-3} 
\multicolumn{1}{|c|}{}                                                                                   & Greedy, 50\%, 75\% & 2.4           \\ \hline
\end{tabu}}
        \end{minipage}
}

\vspace{-3mm}
  \caption{\color{black} {\ourmethod} is more robust than greedy approaches with various thresholds. }
  \vspace{0mm}
  \label{fig: greedy baselines}
\end{figure*}

{
\vspace{1mm}
\color{black}\noindent\textbf{Comparison with greedy approaches.} 
An interesting question is whether a simple greedy policy can effectively address our problem. To explore this possibility, we implemented a series of greedy strategies for comparison. We utilized a per-level detector to detect the workload within a level and set two thresholds, $h_{bottom}$ and $h_{top}$. If the percentage of lookups in the level is less than $h_{bottom}$, the greedy algorithm identifies the workload as write-heavy and increases the compaction policy of the level by one. Conversely, if the percentage of lookups in the level exceeds $h_{top}$, the greedy algorithm recognizes the workload as read-heavy and decreases the compaction policy by one. We consider a wide range of thresholds, including the symmetric thresholds and biased thresholds shown in the right table of Figure~\ref{fig: greedy baselines}, and present four settings in the left figure of Figure~\ref{fig: greedy baselines} for better visualization. The results indicate that some greedy methods can achieve the desired performance for read-heavy or write-heavy workloads. However, for other workloads such as balanced, read-inclined, and write-inclined workloads, they fail to reach the optimal performance. In contrast, {\ourmethod} can self-tune to a near-optimal performance across all workloads, thus providing a relatively more robust performance. The robustness of {\ourmethod} is attributed to its ability to capture the intricate mapping from workload ratio to the optimal policy through a reward mechanism, which is absent in the greedy approaches. 

The right table in Figure~\ref{fig: greedy baselines} summarizes the average performance ranking of all methods, similar to the setting of Table~\ref{table:ranking}. Overall, {\ourmethod} achieves the highest average ranking, surpassing the best setting of a greedy approach we tested. Our findings suggest that while greedy policies may be effective in some scenarios, the use of reinforcement learning methods like {\ourmethod} may provide more robust and comprehensive optimization for LSM-trees.

\begin{figure}[t]
  \centering
    \includegraphics[width=0.66\textwidth]{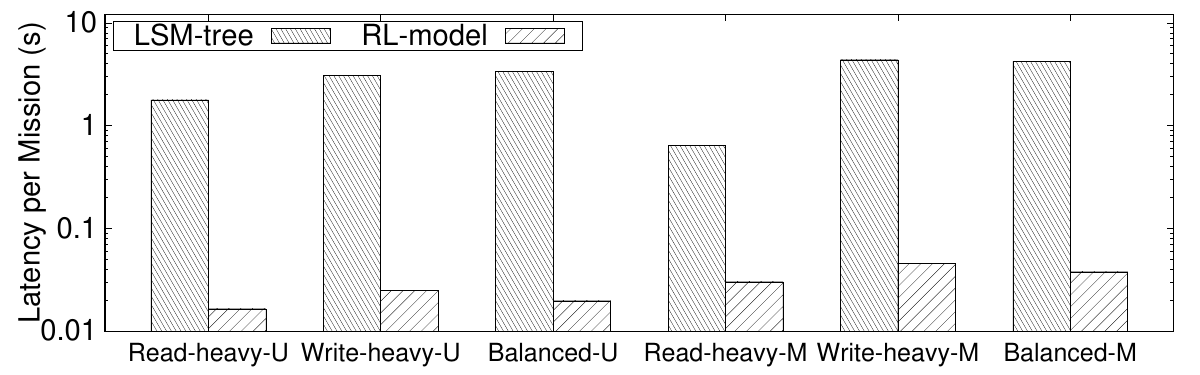}
\vspace{-2mm}
  \caption{\small\color{black} {\ourmethod}'s model update time is relatively insignificant.}
  \label{fig: model updating}
  \vspace{2mm}
\end{figure}

\vspace{1mm}
\noindent\textbf{The model updating time is insignificant.} 
In Figure~\ref{fig: model updating}, we compare the RL model update cost and LSM-tree operation cost per mission on various workloads. ``U'' and ``M'' refer to the uniform Bloom filter scheme and the Monkey scheme, respectively. The Y-axis is in logarithmic scale. Our results show that the model update cost is insignificant compared with the total cost of processing the workload by the LSM-tree. For most workloads, the time spent on updating the model is at most 1\% of the time for processing LSM-tree operations. 
For example, on the balanced workload with the uniform Bloom filter scheme, the average query processing cost per mission is 3.36 seconds, while the average model update cost is only 0.019 seconds.
The small update cost of the RL model of {\ourmethod} is mainly attributed to the reduction of the action space.
}

{\color{black}
\vspace{1mm}
\noindent\textbf{Brute-force learning approaches can be impractical.} 
We evaluated two brute-force training approaches: (1) {\it without a level-based model that reduces the action space}, and  
(2) {\it without the policy propagation and train all levels.} 
We limit the maximum experiment time to 24 hours and use the setting of Figure~\ref{fig: latency no monkey} (c). Our results show that the first approach cannot finish learning within the given time frame while the second approach fails to achieve the optimum policy starting from Level 3 due to insufficient samples. This demonstrates the importance of our optimization strategy. 
}

{
\vspace{1mm}
\color{black}
\noindent
\textbf{Limitations.}
Our work represents a first attempt at using reinforcement learning to optimize LSM-trees, and as such, we have only explored one feasible solution for using machine learning to tune LSM-trees. However, there are opportunities for further improvement of {\ourmethod} by incorporating additional design parameters. For instance, we could learn to adjust the memory allocation for Bloom filters by considering a more refined analysis with non-zero result workloads or adapt size ratios based on a given workload. The challenge here is to maintain a practical action space and a reasonable LSM-tree transition cost. Another limitation of our approach is that the flexible transition degenerates into lazy transition in an extreme read-only case since flexible transition only takes effect when there are updates. It is unclear whether there are better transition methods for this scenario, and this remains an open question for future research. In summary, there is still much to be explored regarding the optimization of LSM-trees using reinforcement learning, and these aspects deserve further study in future works. 
}
\section{Conclusion} 
We present {\ourmethod}, which is an RL-enhanced key-value store that supports dynamic workloads in an online manner.
We integrate into {\ourmethod} a novel {\FLSM} to perform an efficient compaction policy transformation, where a {\FLSM} improves an original LSM-tree in terms of transition costs and delays. Further, we model the system by a level-based RL model {\RLmodel}, which applies DDPG to each level of the {\FLSM} for effective tuning of {\ourmethod} under dynamic workloads. Moreover, we propose a policy propagation method across levels, reducing the demand for training samples.
\vspace{2mm}

\section{Acknowledgement}
This research is supported by NTU SUG-NAP and the Ministry of Education, Singapore, under its AcRF Tier-2 (MOE-T2EP20122-0003). Any opinions, findings and conclusions or recommendations expressed in this paper are those of the author(s) and do not reflect the views of the Ministry of Education, Singapore. We thank Andrew Lim and Weiping Yu for the disucssions during the initial phase of the research.

\bibliographystyle{ACM-Reference-Format}
\bibliography{sample}

\end{document}